\documentclass[aps,prl,reprint,groupedaddress]{revtex4-2}
\usepackage{amsmath}
\usepackage{amssymb}
\usepackage{graphicx}
\usepackage{subfigure}
\usepackage{comment}
\usepackage{ntheorem}
\usepackage{tabularx}
\usepackage{color}
\newcommand{\zb}{\color {black}}
\newcommand{\sx}{\color {black}}
\newcommand{\wsx}{\color {black}}
\begin{document}
\newtheorem{Definition}{Definition}[subsection]
   \title{Constraints of internal symmetry on the non-Hermitian skin effect and bidirectional skin effect under the action of the Hermitian conjugate of time-reversal symmetry}
   \author{Shu-Xuan Wang }
   \email{wangshx65@mail.sysu.edu.cn}
   \affiliation{Guangdong Provincial Key Laboratory of Magnetoelectric Physics and Devices, School of Physics,
   Sun Yat-sen University, Guangzhou 510275, China}

   \date{\today}

   \begin{abstract}
     Non-Hermitian skin effect is a {\zb basic} phenomenon in non-Hermitian {\zb systems}, which means {\zb that an} extensive number of eigenstates can be localized at the boundary. In this {\zb Letter}, we {\zb systematically} investigate {\zb the constraints from all internal symmetries on the} non-Hermitian skin effect in arbitrary dimensions. {\zb By adopting the  powerful} Amoeba formulation, we build a generic correspondence between the various internal symmetries and the behavior of the non-Hermitian skin effect. {\zb Notably}, we find that, {\zb for non-Hermitian systems with the Hermitian conjugate symmetry of time-reversal symmetry~(TRS$^\dagger$)},  the eigenstates can {\zb simultaneously} localize at opposite boundaries,
     which is beyond the Amoeba formulation and we dub the phenomenon {\it bidirectional skin effect}. {\zb Our work provides an overall  perspective from the internal symmetry to the non-Hermitian skin effect.}
   \end{abstract}

   \maketitle

   \emph{Introduction}--- Hermiticity of Hamiltonian is a fundamental assumption for closed system{\zb s}. Once {\zb a} system has gain and loss,
   it can {\zb effectively} be described by a non-Hermitian Hamiltonian\cite{1,2,3,4,5}. In recent years, non-Hermitian systems have
   {\zb attracted very active studies}\cite{5,6,7,8,9,10,11,12,13,14,36}, {\zb accompanying with the discovery of many novel phenomena}, such as exceptional points\cite{5,15,16,17,18,19,20,21,22,23,24,25}, non-Hermitian $PT$ symmetry {\zb breaking}\cite{26,27}, edge burst\cite{28,29,30,31}, etc. {\sx In experiment, such non-Hermitian systems can be realized in photonic crystals\cite{48,49}, ultracold atoms\cite{50,45} and acoustic cavities\cite{51,52}.} {\zb One particularly intriguing phenomenon in non-Hermitian systems is the so-called} non-Hermitian skin effect~(NHSE)\cite{8,10,32,33,34,35,37,38,39,40,41,42,43,44,45,47}, which refers to {\zb the presence of abundant} bulk states at the boundary. {\zb As is known}, {\zb for Hermitian systems  under open boundary conditions~(OBC)} the wave functions of bulk states are always extended, and only topological states are {\zb sharply localized at the boundary}. The spectra of {\zb a} Hermitian system under OBC and periodic boundary condition{\zb s}~(PBC) are {\zb almost} the same except for {\zb the} topological modes. Thus, {\zb the rise of NHSE in a non-Hermitian system} implies that the wave functions and spectra under OBC are {\zb drastically} different from those under PBC. For {\zb one}-dimensional (1D) non-Hermitian system{\zb s}, {\zb both the wave functions and the spectra} of bulk states under OBC can be well described by {\zb the} non-Bloch band theory\cite{10,36}. {\zb For dimensions higher than 1D, building non-Bloch band theory turns out to be quite challenging,
   and only very recently it has been uncovered
   that the Amoeba formulation could provide a general framework for studying higher-dimensional NHSE\cite{44,46,47}.} 
   \par

   Since the Hamiltonian is no longer an Hermitian operator, there are {\zb seven rather than three} internal symmetries for non-Hermitian system{\zb s}\cite{6}. Similar to Hermitian system{\zb s}, {\zb an} internal symmetry will put {\zb certain} constraint on the spectra of non-Hermitian system{s}.
   {\zb Take the particle-hole symmetry as an example}. If the system has an eigenstate with eigenenergy  $E$, {\zb then} there must be another eigenstate {\zb at} $-E$\cite{45}. {\zb Notably}, the internal symmetry also 
   puts constraints on the NHSE. {\zb Some previous works have shown that the 
   TRS$^\dagger$ can result in a class of $\mathbb{Z}_2$ skin effect in 1D non-Hermitian systems\cite{11,34}, while 
   the particle-hole symmetry in arbitrary dimensions enforces two particle-hole partner skin modes to localize at opposite boundaries.
   Although those case-by-case works have revealed the existence of nontrivial interplay between the internal symmetry and the NHSE, the generic
   relation between them remains to be established. }
   \par

   In this {\zb Letter}, we {\zb systematically} investigate the influence of {\zb all seven internal symmetries} on the NHSE through the Amoeba formulation. By {\zb analytically} calculating the transformation of winding number {\zb associated with} the Ronkin function, we {\zb determine} the generic relation between the internal symmetry and the NHSE in arbitrary dimension{\zb s}. {\zb Furthermore, we note that the skin modes 
   in a non-Hermitian system with TRS$^\dagger$ can simultaneously be localized at opposite boundaries, a 
   phenomenon that we dub as {\it bidirectional skin effect}. Notably, we find that this phenomenon is beyond the description of the Amoeba formulation, but can be captured by the TRS$^\dagger$ winding numbers defined by us.}
   \par

   \emph{Internal symmetry of non-Hermitian system{\zb s}}--- {\zb Owing to the lift of the constraint from hermiticity, the type of internal symmetry is enriched in non-Hermitian systems}. The {\zb seven} internal symmetries for non-Hermitian system{\zb s} {\zb include} time-reversal symmetry~(TRS), particle-hole symmetry~(PHS), chiral symmetry~(CS), Hermitian conjugate symmetry of time-reversal symmetry~(TRS$^\dagger$), Hermitian conjugate symmetry of particle-hole symmetry~(PHS$^\dagger$), sublattice symmetry~(SLS) and pseudo Hermitian symmetry\cite{6}. The operators of these internal symmetries can be written {\zb in the} general form,
     \begin{equation}
       \hat{\mathcal{O}} = U_{\mathcal{O}} \hat{R},
       \label{1}
     \end{equation}
   where $U_{\mathcal{O}}$ is a unitary matrix, $\hat{R}$ {\zb the} identity operator for unitary symmetries~(PHS, PHS$^\dagger$, pseudo Hermitian symmetry) and {\zb the complex} conjugate operator for anti-unitary symmetries~(TRS, TRS$^\dagger$, CS, SLS). For a non-Hermitian Hamiltonian, $H(\mathbf{k})$, the transformation {\zb under an internal symmetry operation} can be represented as
     \begin{equation}
      U_{\mathcal{O}}  H(\mathbf{k}) U_{\mathcal{O}}^{-1} = H^{\mathcal{O}}(\mathbf{k}),
      \label{2}
     \end{equation}
   where $U_{\mathcal{O}}$ is the unitary matrix given in Eq.\eqref{1}. For general case{\zb s}, the Hamiltonian $H(\mathbf{k})$ {\zb can be expressed in the general form}
     \begin{equation}
       H(\mathbf{k}) = \sum_{\mathbf{j}} t_{\mathbf{j}} e^{i \mathbf{k} \cdot \mathbf{j}},
       \label{3}
     \end{equation}
   where $\mathbf{j} = (j_1, j_2, \cdots, j_d)$ {\zb denotes} the hopping {\zb vector}, $\mathbf{k} = (k_1, k_2, \cdots, k_d)$ is the momentum vector in dD, {\zb and} $t_{\mathbf{j}}$ is {\zb an} $ s \times s $ hopping matrix if the system which has $s$ {\zb degrees of freedom (like orbital or sublattice) in a unit cell}. {\zb When} the system has {\zb an} internal symmetry, the transformation of the matrix $t_{\mathbf{j}}$ under the {\zb corresponding} internal symmetry {\zb is given by}
     \begin{equation}
       U_{\mathcal{O}}  t_{\mathbf{j}} U_{\mathcal{O}}^{-1} = t_{\mathbf{j}}^{\mathcal{O}}.
       \label{4}
     \end{equation}
    The forms of $H^{\mathcal{O}}(\mathbf{k})$ and $t_{\mathbf{j}}^{\mathcal{O}}$ for specific internal symmetries are given in Supplemental Materials\cite{54}.
   \par

   \emph{Amoeba formulation}--- {\zb For 1D non-Hermitian systems, one can obtain the non-Bloch band theory by directly solving 
   the lattice Schr\"{o}dinger equation. }However, for high{\zb er} dimensional non-Hermitian system{\zb s}, the boundary condition{\zb s} of the Schr\"{o}dinger equation on the lattice {\zb are} much more complicated. {\zb As a consequence}, the Schr\"{o}dinger equation for  higher dimensional non-Hermitian systems {\zb in general} cannot be directly solved,  {\zb and thereby a building of the 
   the higher-dimensional non-Bloch band theory faces great challenging}. {\zb Recently a remarkable progress in solving 
   this fundamental problem is the recognition of the connection between the Amoeba formulation and the NHSE
   in arbitrary dimensions \cite{44}.}
   \par

   {\zb The Ronkin function is a key technique in the Amoeba formulation.} 
   For a non-Hermitian system, whose Hamiltonian under PBC is given by Eq.\eqref{3}, the Ronkin function {\zb characterizing this system is
   of the form}
     \begin{equation}
        R (E, H, \boldsymbol{\mu}) = \int_{T^d} \left(\frac{d k}{2 \pi} \right)^d \ln \det \left[ E - H ( e^{\boldsymbol{\mu} + i \mathbf{k}} ) \right],
        \label{5}
     \end{equation}
   where $T^d = [0,2\pi]^d$ {\zb denotes the dD torus}, and
     \begin{equation}
        \begin{split}
          H ( e^{\boldsymbol{\mu} + i \mathbf{k}} ) & = H(e^{\mu_1 + i k_1}, \cdots , e^{\mu_d + i k_d})  \\
          & = \sum_{\mathbf{j}} t_{\mathbf{j}} e^{\mathbf{j} \cdot (\boldsymbol{\mu} + i \mathbf{k})}
        \end{split}
        \label{6}
     \end{equation}
{\zb with $\boldsymbol{\mu}$ the vector reflecting the localization property of the skin modes. }
Consider a skin mode of the non-Hermitian system,
     \begin{equation}
        \psi_0 (\mathbf{x}) = \beta_1^{x_1} \beta_2^{x_2} \cdots \beta_d^{x_d} \psi,
        \label{7}
     \end{equation}
   where $\mathbf{x} = (x_1, x_2, \cdots, x_d)$ is the position, $\beta_j = e^{\mu_{0,j} + i k_j}$ for the $j$th direction, and the eigenenergy of $\psi_0$ is denoted as $E_0$\footnote{ {\sx Due to OBC, the wave function corresponding to $E_0$ is a linear combination of many modes. However, contributions of some modes vanish in the thermodynamic limit. The rest of the modes have the same vector $\boldsymbol{\mu}$. That means the localization property of these modes are the same. This work only concerns the localization property of the wave function, thus the wave function can be written as the form in Eq.\eqref{7}  }}. According to the Amoeba formulation, the Ronkin function, $R(E_0, H, \boldsymbol{\mu})$, {\zb will take} the minimum value if
     \begin{equation}
       \boldsymbol{\mu} = \boldsymbol{\mu}_0 = (\mu_{0,1}, \cdots, \mu_{0,d}).
       \label{8}
     \end{equation}
   In other words, if $E_0$ belongs to the spectrum of the non-Hermitian system under OBC, and the eigenstate corresponding to it is $\psi_0$ given in Eq.\eqref{7}, {\zb then} the winding number
     \begin{equation}
        \begin{split}
          w_m & = \frac{1}{2 \pi i} \int_{0}^{2 \pi} d k_m \partial_{k_m} \ln \det \left[ E - H ( e^{\boldsymbol{\mu}_0 + i \mathbf{k}} ) \right] \\
          & = 0 \quad \mathrm{or} \quad  \mathrm{ill \ defined}
        \end{split}
        \label{9}
     \end{equation}
   for $\forall m = 1,2, \cdots, d$\cite{44}\footnote{According to Ref.\cite{44}, the Ronkin function is convex in the entire $\boldsymbol{\mu}$ space, thus Eq.\eqref{9} is the condition for the minimum value of Ronkin function. The Amoeba formulation also point out that if $E$ belongs to the OBC spectrum of the non-Hermitian system and the wave function corresponding to $E$ has the decaying factor, $\boldsymbol{\mu}_0$, then there existss a vector, $\mathbf{k}_0$, such that $E$ is an eigenvalue of $H ( e^{\boldsymbol{\mu}_0 + i \mathbf{k}_0 } )$. Thus for specific value of $\mathbf{k}$, the winding number in Eq.\eqref{9} is ill defined. }{\sx (the derivation for Eq.\eqref{9} is given in Supplemental Materials\cite{54})}.
   \par

   \emph{Constraints of internal symmetry on the NHSE}--- Now, we consider a non-Hermitian system, whose Hamiltonian under PBC is given by Eq.\eqref{3}, and has an internal symmetry denoted by $\mathcal{O}$. Assume that this system has a skin mode with eigenenergy $E$ and decay{\zb ing factor}
     \begin{equation}
        \boldsymbol{\mu} = (\mu_1, \cdots, \mu_d).
        \label{10}
     \end{equation}
   We further define a winding number of the form
     \begin{equation}
        w_m^{\mathcal{O}}  = \frac{1}{2 \pi i} \int_{0}^{2 \pi} d k_m \partial_{k_m} \ln \det \left[ U_{\mathcal{O}} \left(E - H ( e^{\boldsymbol{\mu} + i \mathbf{k}} ) \right) U_{\mathcal{O}}^{-1} \right].
        \label{11}
     \end{equation}
    {\zb By a straightforward derivation (see more details in the Supplemental Materials\cite{54}), we find that} 
     \begin{equation}
        w_m^{\mathcal{O}} = \pm \frac{1}{2 \pi i} \int_{0}^{2 \pi} d k_m \partial_{k_m} \ln \det \left[ \tilde{E}_{\mathcal{O}} - H ( e^{\tilde{\boldsymbol{\mu}}_{\mathcal{O}} + i \mathbf{k}} ) \right],
        \label{12}
     \end{equation}
   where $\tilde{E}_{\mathcal{O}} = \pm E \ \mathrm{or} \pm E^{*}$, and $\tilde{\boldsymbol{\mu}}_{\mathcal{O}} = \pm \boldsymbol{\mu}$ for different internal symmetries, as shown in  {\zb Table} \ref{table1}.
     \begin{table}
       \begin{tabularx}{250pt}{|>{\centering}X|>{\centering}X|>{\centering}X|>{\centering}X|>{\centering}X|>{\centering}X|>{\centering}X|>{\centering}X|}
         \toprule
         $\mathcal{O}$ & $\mathcal{T}_{+}$ & $\mathcal{C}_{-}$ & $\mathcal{S}$ & $\mathcal{C}_{+}$ & $\mathcal{T}_{-}$ & $\Gamma$ & $\eta$
         \tabularnewline[2pt]
         \hline
         $\tilde{E}_{\mathcal{O}}$   & $E^{*}$  &  $-E$  & $-E^{*}$  &  $E$  &  $-E^{*}$  &  $-E$  &  $E^{*}$
         \tabularnewline[2pt]
         \hline
         $\tilde{\boldsymbol{\mu}}_{\mathcal{O}}$ & $\boldsymbol{\mu}$ & $-\boldsymbol{\mu}$ & $-\boldsymbol{\mu}$ & $-\boldsymbol{\mu}$ & $\boldsymbol{\mu}$ & $\boldsymbol{\mu}$ & $-\boldsymbol{\mu}$
         \tabularnewline[2pt]
         \botrule
       \end{tabularx}
       \caption{The eigenenergy and decaying factor of the skin mode related to Eq.\eqref{12} for different internal symmetry. The first row is the internal symmetry, the second row is the value of $\tilde{E}_{\mathcal{O}}$, and the third row is the value of $\tilde{\boldsymbol{\mu}}_{\mathcal{O}}$ in Eq.\eqref{12} under the corresponding internal symmetry.  $\mathcal{T}_{+}$, $\mathcal{C}_{-}$, $\mathcal{S}$, $\mathcal{C}_{+}$, $\mathcal{T}_{-}$, $\Gamma$ and $\mathcal{\eta}$ represent TRS, PHS, CS, TRS$^\dagger$, PHS$^\dagger$, SLS and pseudo Hermitian symmetry, respectively.}
       \label{table1}
     \end{table}
     \par

    {\zb Next, using the fact that the determinant of a matrix does not change under a unitary transformation, i.e.,} 
       \begin{equation}
         \det \left[ U_{\mathcal{O}} \left(E - H  \right) U_{\mathcal{O}}^{-1} \right] = \det \left[ (E - H )  \right],
         \label{13}
       \end{equation}
      {\zb it is readily found that} 
        \begin{equation}
         w_m^{\mathcal{O}} = \frac{1}{2 \pi i} \int_{0}^{2 \pi} d k_m \partial_{k_m} \ln \det \left[ E - H ( e^{\boldsymbol{\mu} + i \mathbf{k}} ) \right].
         \label{14}
        \end{equation}
    {\zb As assumed in the beginning that this system has a skin mode with eigenenergy $E$ and decaying factor $\boldsymbol{\mu}$, from 
    the discussion around Eq.\eqref{7} we know that }
    \begin{equation}
          w_m^{\mathcal{O}} = 0 \quad \mathrm{or} \quad  \mathrm{ill \ defined},
          \label{15}
        \end{equation}
    for $\forall m = 1,2,\cdots, d$. {\zb Applying this result back into Eq.\eqref{12}, one can 
    immediately reach the conclusion that if there exists a skin mode at $E$ and 
    a given boundary determined by the decaying factor $\boldsymbol{\mu}$, then there 
    must exist another symmetry-enforced skin mode at $\tilde{E}_{\mathcal{O}}$ and the same place (for $\tilde{\boldsymbol{\mu}}_{\mathcal{O}}=\boldsymbol{\mu}$) or opposite boundaries (for
   $\tilde{\boldsymbol{\mu}}_{\mathcal{O}}=-\boldsymbol{\mu}$). The constraints of the seven internal 
   symmetries on the NHSE summarized in Table \ref{table1} is one of the central results of this work. }
   
   {\zb Before proceeding, we would like to give a remark on the result shown in Table \ref{table1}. From 
   the third column of Table \ref{table1}, one can see that the particle-hole symmetry
   enforces two particle-hole partner skin modes to localize at opposite boundaries. As aforementioned, this generic result has 
   already been obtained in a previous work\cite{45}, however, the approach therein is much more involved and specific 
   for the particle-hole symmetry. Thanks to the powerfulness of the Amoeba formulation for describing NHSE, 
   here we have determined the complete picture in a rather neat way. }
   \par

      {\it  Example illustration.---} To show the correctness of the analytical theory,  we consider a 2D non-Hermitian system for illustration.  The concrete Hamiltonian under PBC is given by
        \begin{equation}
          H_{\mathcal{T}_{+}} (k_x,k_y) =
          \begin{pmatrix}
            t_{-1} e^{- i k_x} + t_1 e^{ i k_x}  &  c + p_{-1} e^{- i k_y}  \\
            c + p_{1} e^{ i k_y}    &   w_{-1} e^{- i k_x} + w_1 e^{ i k_x}
          \end{pmatrix},
          \label{16}
        \end{equation}
        where $t_1$, $t_{-1}$, $p_1$, $p_{-1}$, $w_1$, $w_{-1}$ {\zb and $c$} are real numbers.
      It is {\zb easy to check} that the above Hamiltonian has TRS, and the TRS operator is given by 
        \begin{equation}
          \hat{\mathcal{T}}_{+} = \mathbf{I}_{2 \times 2} \mathcal{K},
          \label{17}
        \end{equation}
      where {\zb $\mathbf{I}_{2 \times 2}$} is the $2 \times 2$ identity matrix and {\zb $\mathcal{K}$ is the complex conjugate operator}.
      \par
\begin{figure*}
        \centering
            \subfigure[]{\includegraphics[scale=0.45]{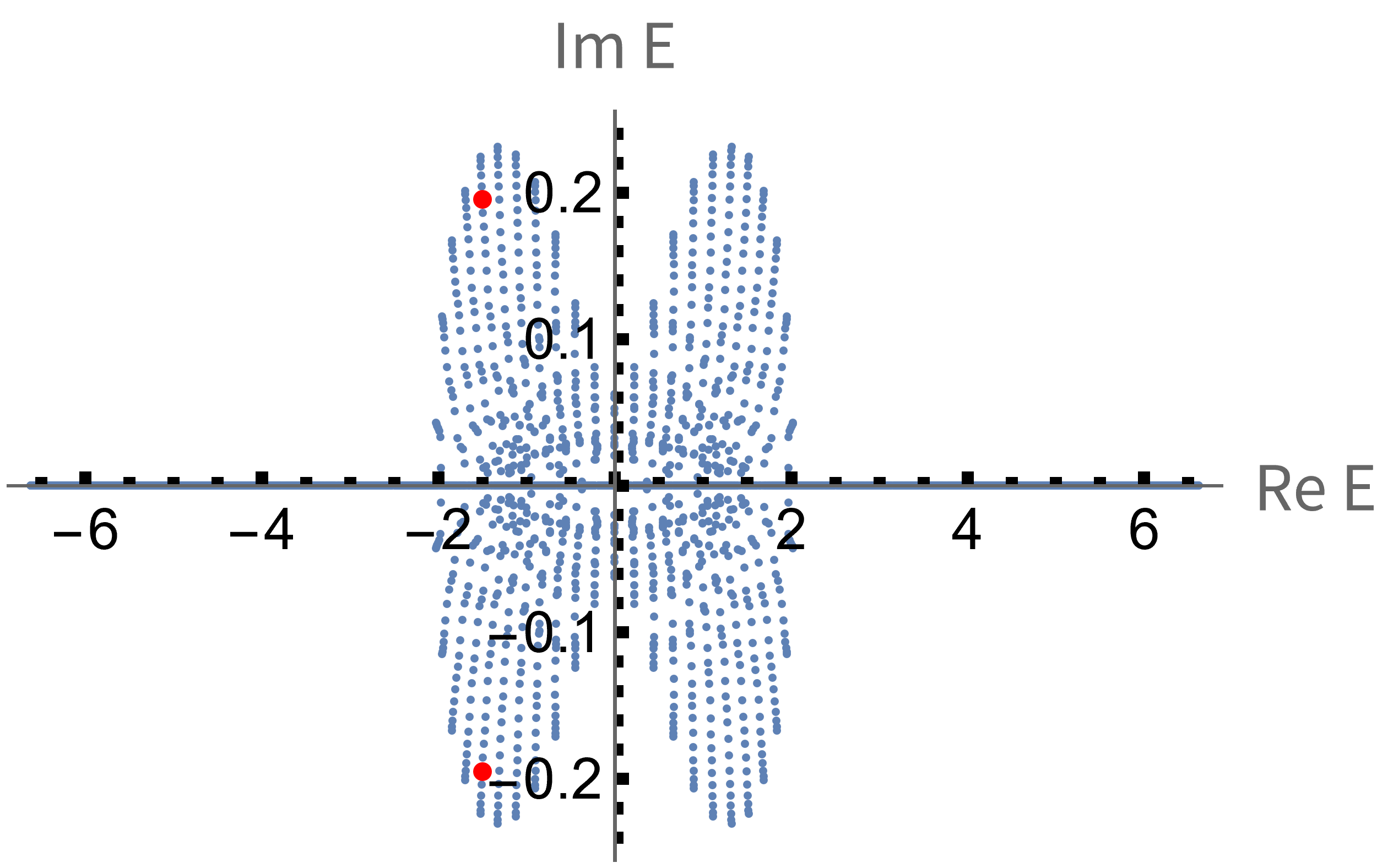} \label{fig1a}}
            \quad
            \subfigure[]{\includegraphics[scale=0.3]{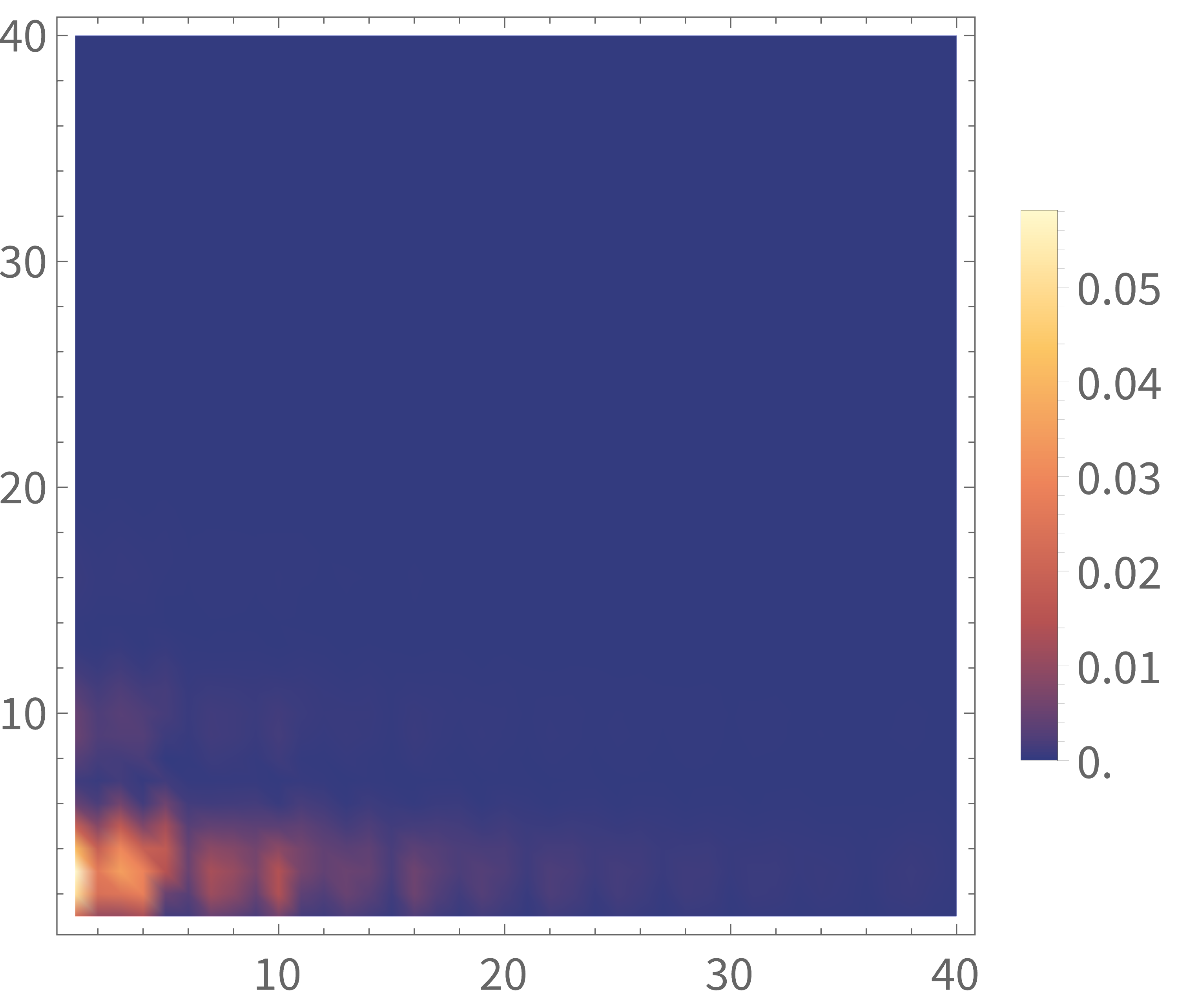} \label{fig1b}}
            \quad
            \subfigure[]{\includegraphics[scale=0.3]{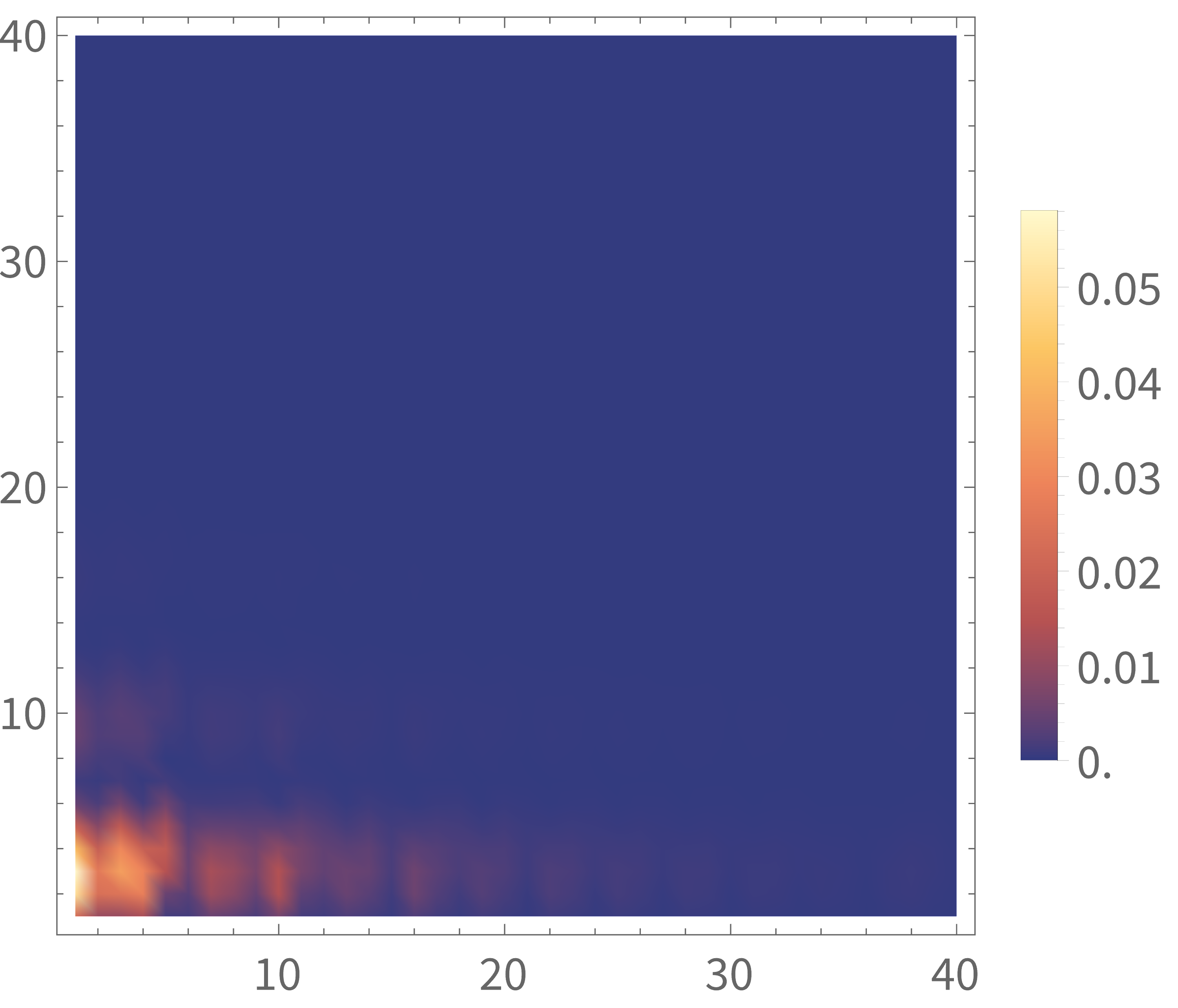} \label{fig1c}}
            \caption{(a)The blue dots refer to eigenenergies of the system given by Eq.\eqref{16} under OBC. The lower red dot corresponds to $E_{TRS1} = -1.5 - 0.195 i$, and the upper one corresponds to $E_{TRS2} = -1.5 + 0.195 i$. (b)The probability density profile corresponding to $E_{TRS1}$. (c)The probability density profile corresponding to $E_{TRS2}$. \  Values of parameters are $t_1=2$, $t_{-1}=1$, $w_1=1.5$, $w_{-1}=3.3$, $p_1=1.8$, $p_{-1}=2.6$, and $c=0.5$. The system size  is $40\times40$.}
            \label{fig1}
      \end{figure*}

      {\zb The spectra of this Hamiltonian under OBC for a specific set of parameter values are given in Fig.\ref{fig1}(a). Consider 
      two eigenstates related by TRS. Without loss of generality, we choose the pair with  $E_{TRS1} =E_{TRS2}^{*}= -1.5 - 0.195 i$
      that are belonged to the OBC spectra. As shown in Figs.\ref{fig1}(b) and \ref{fig1}(c), the probability density profile of 
      these two skin modes are localized at the same boundary, agreeing with the prediction summarized in the second column of Table \ref{table1}.}
      Examples for other internal symmetries are {\zb provided} in {\zb the} Supplemental Materials\cite{54}, and all numerical results agree with 
      the predictions summarized in Table \ref{table1}.
      \par

\begin{figure}
          \subfigure[]{\includegraphics[scale=0.45]{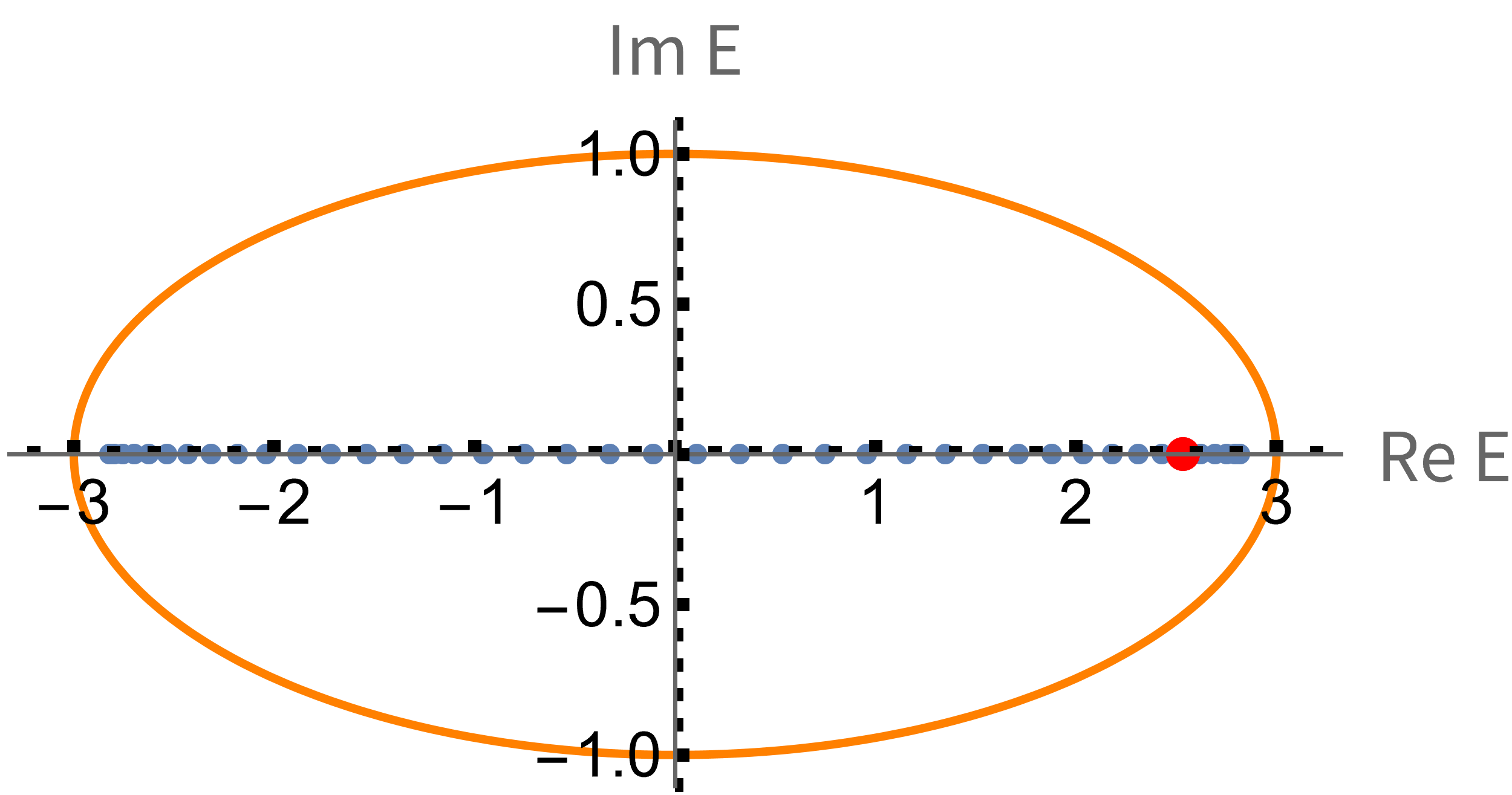} \label{fig2a}}
          \quad
          \subfigure[]{\includegraphics[scale=0.4]{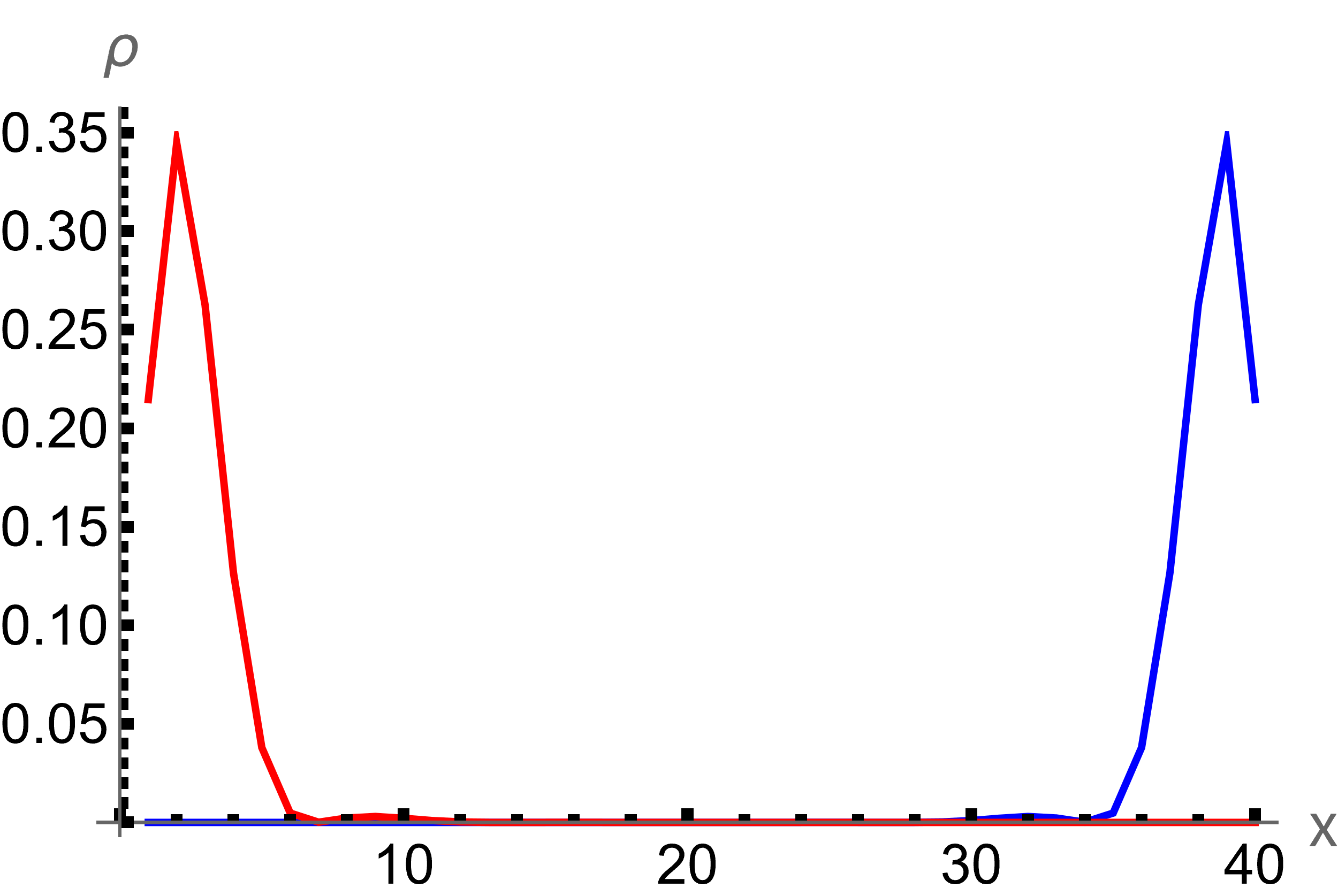} \label{fig2b}}

          \caption{(a)The blue dots refer to eigenenergies of the system given by Eq.\eqref{18} under OBC for the case that $\gamma = 0$, and the orange loop corresponds to the PBC spectrum for $\gamma = 0$. The two bands under PBC are degenerate, and both of them correspond to the same orange loop in the complex plane. The red dot corresponds to $E_{TRS^{\dagger}1} = 2.53$ . (b)The blue and red lines are the distributions of the two eigenstates corresponding to $E_{TRS^{\dagger}1}$. \  Values of other parameters are $t_1=1$, $t_{-1}=2$, and the system size is $40$.}
          \label{fig2}
        \end{figure}

\begin{figure}
          \subfigure[]{\includegraphics[scale=0.45]{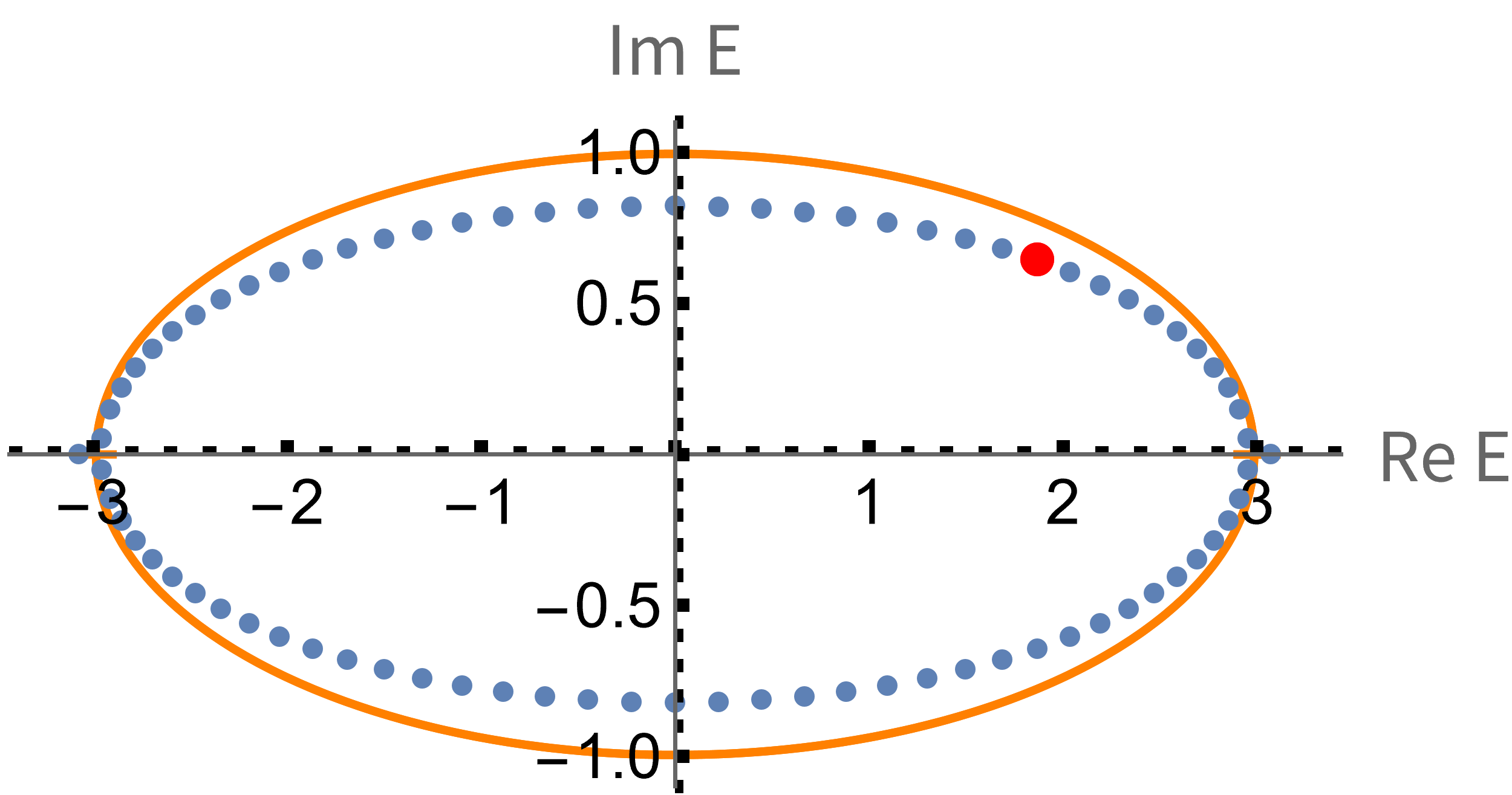} \label{fig3a}}
          \quad
          \subfigure[]{\includegraphics[scale=0.4]{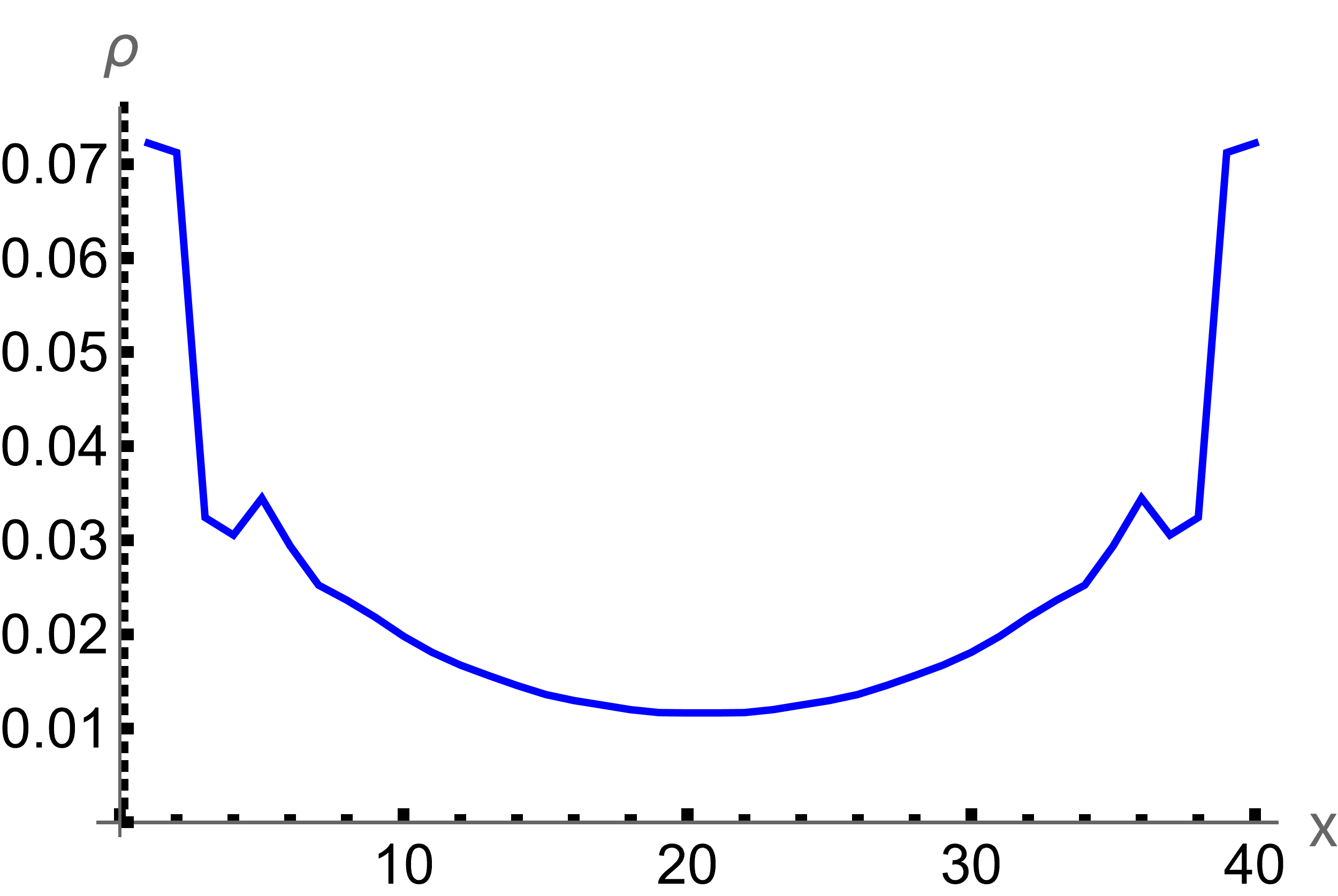} \label{fig3b}}

          \caption{(a)The blue dots are the locations of the eigenenergies of the system given by Eq.\eqref{18} under OBC for the case with $\gamma = 0.1$, and the orange loop corresponds to the PBC spectrum for $\gamma = 0.1$. The two bands under PBC are also degenerate, and both of them correspond to the same orange loop in the complex plane. The red dot is $E_{TRS^{\dagger}2} = 1.87+0.64i$. (b)The blue line is the distribution of the eigenstate corresponding to $E_{TRS^{\dagger}2}$. \  Values of other parameters are $t_1=1$, $t_{-1}=2$, and the system size is $40$.}
          \label{fig3}
        \end{figure}

      \emph{Bidirectional skin effect.---} From Table \ref{table1}, one can see that the fifth column corresponding to TRS$^\dagger$ is quite 
      special since only for this symmetry $\tilde{E}_{\mathcal{O}}=E$. The invariance of $E$  and the change of $\boldsymbol{\mu}$ to $-\boldsymbol{\mu}$ under the TRS$^\dagger$ operation indicate that, under OBC, there will be two degenerate skin modes localizing at opposite boundaries of the system. However, we find that such degenerate skin modes  are fragile. For general cases, 
      the degeneracy disappears and the eigenstate is extended or simultaneously localized at two opposite boundaries. To illustrate 
      this fact, we construct a simple 1D model of the form



        \begin{equation}
            H(k) =
            \begin{pmatrix}
              t_{-} e^{-i k} + t_{+} e^{i k}  &
              \gamma
              \\
              \gamma  &
              t_{-} e^{i k} + t_{+} e^{-i k}
            \end{pmatrix}.
            \label{18}
        \end{equation}
      The TRS$^\dagger$ operator for this system is
        \begin{equation}
          \hat{\mathcal{C}}_{+} = U_{\mathcal{C}_{+}} \mathcal{K}  =
          \begin{pmatrix}
            0  &  1  \\
            1  &  0
          \end{pmatrix}
          \mathcal{K}.
          \label{19}
        \end{equation}
      In the limit of $\gamma = 0$, the OBC spectra of this system are doubly degenerate, and two eigenstates corresponding to the same eigenenergy, $E$ are localized at opposite boundaries~(Fig.\ref{fig2}). Once $\gamma \not= 0$, the double degeneracy of the OBC spectra is lifted, and  we find eigenstates that are simultaneously localized at two opposite boundaries~(Fig.\ref{fig3}). 

      \par
       Remarkably, we find that the above phenomenon is in fact beyond the description of Amoeba formulation.
       To see this, let us return to the Amoeba formulation. Assuming that $E$ belongs to the spectra of the system under OBC, the winding number about $E$ for the Hamiltonian under PBC is
        \begin{equation}
           w_m^{PBC} = \frac{1}{2 \pi i} \int_{0}^{2 \pi} d k_m \partial_{k_m} \ln \det \left[ E - H(\mathbf{k}) \right],
           \label{20}
        \end{equation}
      where $m=1,2,\cdots,d$.
      After the transformation of $U_{\mathcal{C}_{+}}$, we find that~(see Supplemental Materials for details\cite{54})
        \begin{equation}
           \begin{split}
              w_m^{PBC} & = \frac{1}{2 \pi i} \int_{0}^{2 \pi} d k_m \partial_{k_m} \ln \det \left[ U_{\mathcal{C}_{+}} \left( E - H(\mathbf{k}) \right) U_{\mathcal{C}_{+}}^{-1} \right]
              \\
              & = \frac{1}{2 \pi i} \int_{0}^{2 \pi} d k_m \partial_{k_m} \ln \det \left[ E - H(-\mathbf{k})^{T} \right]
              \\
              & = - w_m^{PBC}.
           \end{split}
           \label{21}
        \end{equation}
      One immediately obtains 
        \begin{equation}
           w_m^{PBC} = \ 0 \quad \mathrm{or} \quad \mathrm{ill} \  \mathrm{defined},
           \label{22}
        \end{equation}
      for $\forall m=1,2,\cdots,d$.  Since
        \begin{equation}
           H(\mathbf{k}) = H(e^{\boldsymbol{\mu} + i \mathbf{k}}) |_{\boldsymbol{\mu} = \mathbf{0}},
           \label{23}
        \end{equation}
      according to the Amoeba formulation, all eigenstates of this system are extended except for topological modes, which means that the NHSE does not exist in this system with TRS$^\dagger$. However, as shown above, we find that different from traditional skin modes, eigenstates can simultaneously localize at opposite boundaries. For the convenience of describing this phenomenon, we term it {\it bidirectional skin effect}. It may be worth emphasizing that here our findings can be  applied to arbitrary dimensions. In 1D, the bidirectional skin effect is just reduced to the $\mathbb{Z}_2$ skin effect discovered in Ref.\cite{34}.
      \par

      In the following,  we develop a topological 
      invariant to capture the bidirectional skin effect. The steps are as follows. First, 
      we note that the winding number in Eq.\eqref{20} can be rewritten as
        \begin{equation}
           \begin{split}
             w_m^{PBC} (E) & = \frac{1}{2 \pi i} \int_{0}^{2 \pi} d k_m \partial_{k_m} \ln
              \left[ \prod_{n=1}^{2N} E - E_n (\mathbf{k}) \right]
              \\
              & = \sum_{n=1}^{2N} \frac{1}{2 \pi i} \int_{0}^{2 \pi} d k_m \partial_{k_m} \ln
              \left[  E - E_n (\mathbf{k}) \right]
              \\
              & = \sum_{n=1}^{2N} w_m^{(n)} (E),
           \end{split}
           \label{24}
        \end{equation}
      where $E_n (\mathbf{k})$ is the $n$th band of the system under PBC, and
        \begin{equation}
           w_m^{(n)} (E) = \frac{1}{2 \pi i} \int_{0}^{2 \pi} d k_m \partial_{k_m} \ln
           \left[  E - E_n (\mathbf{k}) \right],
           \label{25}
        \end{equation}
      is the winding number {\zb of the} $n$th band. We find that if {\zb under OBC the eigenstate at $E$}  is extended, {\zb then}
        \begin{equation}
           w_m^{(n)} (E) = 0
           \label{26}
        \end{equation}
      for $\forall m = 1,2,\cdots, d,\ \mathrm{and} \ n=1,2,\cdots,2N$. {\zb In contrast}, if the eigenstate is simultaneously
      localized at opposite boundaries, {\zb pairs of non-zero winding numbers} will appear\footnote{For general cases, the band, $E_{\mu} (\mathbf{k})$, is multi-valued. Here, we always choose the branch that makes $E_{\mu} (\mathbf{k})$ be a closed loop but not an open curve in the complex plane.}, i.e.,
        \begin{equation}
           w_m^{(p)} (E) = - w_m^{(q)} (E) \not= 0,
           \label{27}
        \end{equation}
      where {\zb the} $q$th band is related to the $p$th band by TRS$^\dagger$, i.e.,
        \begin{equation}
          E_q (\mathbf{k}) = \hat{\mathcal{C}}_{+} E_p (\mathbf{k}).
          \label{28}
        \end{equation}
      Therefore, for a non-Hermitian system with TRS$^\dagger$, whether the eigenstate at a given $E$ is a bidirectional skin mode or not can be described by
        \begin{equation}
           \nu_m (E) = \frac{1}{2} \sum_{n=0}^{N-1} |w_m^{(2n+1)} (E) - w_m^{(2n+2)} (E)|,
           \label{29}
        \end{equation}
      where $m=1,2,\cdots,d$. {\zb Here the $(2n+1)$th and $(2n+2)$th bands} are related by TRS$^\dagger$, and the system has $2N$ bands {\zb in total} \footnote{If the number of bands is odd, there exists a band, $E_{r} (\mathbf{k})$, such that $E_r (\mathbf{k}) = \hat{\mathcal{C}}_{+} E_r (\mathbf{k})$. The winding number of this band, $w_m^{(r)} (E)$ must be zero and gives no contribution to the topological invariant, $\nu_m (E)$. Hence, we only consider the case that the system has an even number of bands }. We call these topological invariants TRS$^\dagger$ winding numbers. If $\nu_m (E) = 0$ for $\forall m = 1,2,\cdots, d$, {\zb the eigenstate at $E$} is extended in the bulk. In contrast, if $\nu_m (E) \not= 0$  for one or more $\nu_m (E)$, the eigenstate at $E$ is a bidirectional skin mode. See the spectrum shown in Fig.\ref{fig3a}, it is not {\zb hard} to find that for the bidirectional skin mode corresponding to $E_{TRS^{\dagger}2}$, $\nu (E_{TRS^{\dagger}2}) = 1$\footnote{For the case that $\gamma = 0$ shown in Fig.\ref{fig2}, it is also not hard to find that $\nu (E_{TRS^{\dagger}1}) = 1$. On the other hand, two degenerate skin modes corresponding to $E_{TRS^{\dagger}1}$ can consist of two states, which are localized at opposite boundaries. Thus, this case can be seen as a special case of bidirectional skin effect.}. Examples for bidirectional skin effect in 2D non-Hermitian systems are given in Supplemental Materials\cite{54}.
      \par

      \emph{Conclusion and discussion}--- In this Letter, we have systematically investigated the behavior of NHSE under various internal symmetries based on the Amoeba formulation and obtained their constraints on the NHSE. We find that for a non-Hermitian system with an internal symmetry $\mathcal{O}$, if it has a skin mode with a decaying factor $\boldsymbol{\mu}$ and at eigenenergy $E$, then there must exist another skin mode at  eigenenergy $\tilde{E}_{\mathcal{O}}$, which localizes at the same~(opposite) boundary of the original skin mode if $\tilde{\boldsymbol{\mu}}_{\mathcal{O}} = \boldsymbol{\mu}$~($\tilde{\boldsymbol{\mu}}_{\mathcal{O}} = -\boldsymbol{\mu}$). Furthermore, we have found the bidirectional skin effect in non-Hermitian systems with TRS$^\dagger$, and we have introduced
      a topological invariant named as TRS$^\dagger$ winding numbers to describe this phenomenon. {\sx From the aspect of 38-fold symmetry classes, for a non-Hermitian system with symmetry class $\mathcal{Q}$, if $\mathcal{Q}$ contains TRS$^\dagger$, the system can have bidirectional skin effect, otherwise, only the conventional skin effect is allowed.} To conclude, 
      we have established the overall picture of the relation between the internal symmetry and the NHSE. As our findings neither rely on the dimension nor specific non-Hermitian Hamiltonians, they admit a wide application. {\wsx In experiment, the ultracold-atom systems can be utilized as the platform to observe the novel behaviors of non-Hermitian skin effect under internal symmetries\cite{50,45,53}. Furthermore, our formula can be used to research the properties of non-Hermitian skin effect under crystalline symmetries and this is a focus in our future work. }
      \par

      \emph{Acknowledgements}--- The author thanks Zhongbo Yan for helpful suggestions.  This work is supported by the National Natural Science Foundation of China~(Grant No.12174455) and the Natural Science Foundation of Guangdong Province~(Grant No.2021B1515020026).

   \bibliography{internalsymmetry}

   \clearpage

   \onecolumngrid
    \begin{center}
     $ \boldsymbol{\mathbf{Supplemental \ Material}}$
    \end{center}
 
   \renewcommand\thesection{\Roman{section}}

   \section{Transformation of Hamiltonian Under Internal Symmetries}
     In this section, we give the specific form of $ H^{\mathcal{O}}(\mathbf{k})$ and $t_{\mathbf{j}}^{\mathcal{O}}$ in Eq.(2) and Eq.(4) in main text.
     \par

     According to Ref.\cite{6}, for the non-Hermitian Hamiltonian, $H(\mathbf{k})$, the transformation of it under internal symmetries can be represented as
       \begin{gather}
        U_{\mathcal{T}_{+}}  H(\mathbf{k}) U_{\mathcal{T}_{+}}^{-1} = H^{*}(-\mathbf{k}), 
        \label{S1} 
        \tag{S1}
        \\
        U_{\mathcal{C}_{-}}  H(\mathbf{k}) U_{\mathcal{C}_{-}}^{-1} = -H^{T}(-\mathbf{k}), 
        \label{S2} 
        \tag{S2}
        \\
        U_{\mathcal{S}} H(\mathbf{k}) U_{\mathcal{S}}^{-1} = - H^{\dagger} (\mathbf{k}),
        \label{S3} 
        \tag{S3}
        \\
        U_{\mathcal{C}_{+}}  H(\mathbf{k}) U_{\mathcal{C}_{+}}^{-1} = H^{T}(-\mathbf{k}),
        \label{S4} 
        \tag{S4}
        \\
        U_{\mathcal{T}_{-}}  H(\mathbf{k}) U_{\mathcal{T}_{-}}^{-1} = - H^{*}(-\mathbf{k}),
        \label{S5} 
        \tag{S5}
        \\
        U_{\Gamma} H(\mathbf{k}) U_{\Gamma}^{-1} = - H (\mathbf{k}),
        \label{S6} 
        \tag{S6}
        \\
        U_{\mathcal{\eta}} H(\mathbf{k}) U_{\mathcal{\eta}}^{-1} = H^{\dagger} (\mathbf{k}),
        \label{S7} 
        \tag{S7}
       \end{gather}
      where $U_{\mathcal{T}_{+}}$, $U_{\mathcal{C}_{-}}$, $U_{\mathcal{S}}$, $U_{\mathcal{C}_{+}}$, $U_{\mathcal{T}_{-}}$, $U_{\Gamma}$ and $U_{\mathcal{\eta}}$ are unitary matrix of TRS, PHS, CS, TRS$^\dagger$, PHS$^\dagger$, SLS and pseudo Hermitian symmetry respectively. Substituting $H(\mathbf{k})$ in Eq.(3) into equations above, we obtain the the transformation of matrix $t_{\mathbf{j}}$ under the internal symmetries,
        \begin{gather}
          U_{\mathcal{T}_{+}}  t_{\mathbf{j}} U_{\mathcal{T}_{+}}^{-1} = t_{\mathbf{j}}^{*}, 
          \label{S8} 
           \tag{S8} 
          \\
          U_{\mathcal{C}_{-}} t_{\mathbf{j}} U_{\mathcal{C}_{-}}^{-1} = -t_{-\mathbf{j}}^{T}, 
          \label{S9} 
          \tag{S9}
          \\
          U_{\mathcal{S}} t_{\mathbf{j}} U_{\mathcal{S}}^{-1} = -t_{-\mathbf{j}}^{\dagger},
          \label{S10} 
          \tag{S10}
          \\
          U_{\mathcal{C}_{+}}  t_{\mathbf{j}} U_{\mathcal{C}_{+}}^{-1} = t_{-\mathbf{j}}^{T},
          \label{S11} 
          \tag{S11}
          \\
          U_{\mathcal{T}_{-}}  t_{\mathbf{j}} U_{\mathcal{T}_{-}}^{-1} = -t_{\mathbf{j}}^{*},
          \label{S12} 
          \tag{S12}
          \\
          U_{\Gamma} t_{\mathbf{j}} U_{\Gamma}^{-1} = - t_{\mathbf{j}},
          \label{S13} 
          \tag{S13}
          \\
          U_{\mathcal{\eta}} t_{\mathbf{j}} U_{\mathcal{\eta}}^{-1} = t_{-\mathbf{j}}^{\dagger}.
          \label{S14} 
          \tag{S14}
        \end{gather}

  \section{Derviation for Eq.(9)}
      The Ronkin function about the non-Hermitian system is 
        \begin{equation}
          R (E, H, \boldsymbol{\mu}) = \int_{T^d} \left(\frac{d k}{2 \pi} \right)^d \ln \det \left[ E - H ( e^{\boldsymbol{\mu} + i \mathbf{k}} ) \right]
          \label{S15}
          \tag{S15}
        \end{equation}
      If $R (E, H, \boldsymbol{\mu})$ takes the minimum value at
        \begin{equation}
          \boldsymbol{\mu} = \boldsymbol{\mu}_0 = (\mu_{0,1}, \cdots, \mu_{0,d}),
          \label{S16}
          \tag{S16}
        \end{equation}
      $R (E, H, \boldsymbol{\mu})$ satisfies
        \begin{equation}
          \left. \frac{\partial R (E, H, \boldsymbol{\mu})}{\partial \mu_j} \right|_{\boldsymbol{\mu} = \boldsymbol{\mu}_0} = \left. \int_{T^d} \left(\frac{d k}{2 \pi} \right)^d \frac{\partial}{\partial \mu_j}  \ln \det \left[ E - H ( e^{\boldsymbol{\mu} + i \mathbf{k}} ) \right] \right|_{\boldsymbol{\mu} = \boldsymbol{\mu}_0}  = 0
          \label{S17}
          \tag{S17}
        \end{equation}
      for $j = 1,2,\cdots,d$. Since 
        \begin{equation}
          H ( e^{\boldsymbol{\mu} + i \mathbf{k}} )  = H(e^{\mu_1 + i k_1}, \cdots , e^{\mu_d + i k_d}) ,
          \label{S18}
          \tag{S18}
        \end{equation}
      we can replace $\frac{\partial}{\partial \mu_i}$ with $- i \frac{\partial}{\partial k_j}$ in Eq.\eqref{S17}. After that, Eq.\eqref{S17} can be written as 
        \begin{equation}
          \left. \frac{\partial R (E, H, \boldsymbol{\mu})}{\partial \mu_j} \right|_{\boldsymbol{\mu} = \boldsymbol{\mu}_0} =-i \int_{T^d} \left(\frac{d k}{2 \pi} \right)^d \frac{\partial}{\partial k_j}  \ln \det \left[ E - H ( e^{\boldsymbol{\mu}_0 + i \mathbf{k}} ) \right] = 0.
          \label{S19}
          \tag{S19}
        \end{equation}
      The value of the winding number
        \begin{equation}
          w_j  = \frac{1}{2 \pi i} \int_{0}^{2 \pi} d k_j \frac{\partial}{\partial k_j} \ln \det \left[ E - H ( e^{\boldsymbol{\mu}_0 + i \mathbf{k}} ) \right]
          \label{S20}
          \tag{S20}
        \end{equation}
      is a constant. Hence,
        \begin{equation}
          -i \int_{T^d} \left(\frac{d k}{2 \pi} \right)^d \frac{\partial}{\partial k_j}  \ln \det \left[ E - H ( e^{\boldsymbol{\mu}_0 + i \mathbf{k}} ) \right] = \int_{T^{d-1}} \left(\frac{d k}{2 \pi} \right)^{d-1} w_j = w_j.
          \label{S21}
          \tag{S21}
        \end{equation}
      Thus, the condition for the minimum value of Ronkin function is
        \begin{equation}
          w_j =  0 \quad \mathrm{or} \quad  \mathrm{ill \ defined}
          \label{S22}
          \tag{S22}
        \end{equation}
      for $\forall m = 1,2,\cdots, d$, i.e. Eq.(9) in the main text.

   \section{Results in Eq.(12) and table $\mathbf{I}$}
       In this section, we give the demonstrations about results given in Eq.(12) and table $\mathbf{I}$ in the main text.

     \subsection{Properties about the expression of winding number}
       Consider a $s\times s$ matrix, $A(k)$, which depend on the parameter, $k$. Then, the winding number of $A$ about $E$ is 
         \begin{equation}
            w (E) = \frac{1}{2 \pi i} \int_{0}^{2 \pi} d k \partial_k \ln \det \left[ E - A(k) \right].
            \tag{S23}
            \label{S23}
         \end{equation}
       There are 4 useful properties about the winding number for the following demonstrations,
         \begin{equation}
            \begin{split}
                w_{A\mathit{1}} (E) & = \frac{1}{2 \pi i} \int_{0}^{2 \pi} d k \partial_k \ln \det \left[ E - A(-k) \right]
                \\
                & = \frac{1}{2 \pi i} \int_{2 \pi}^{0} d (-k) \partial_{-k} \ln \det \left[ E - A(-k) \right]
                \\
                & = \frac{-1}{2 \pi i} \int_{0}^{2 \pi} d k \partial_k \ln \det \left[ E - A(k) \right]
                \\
                & = - w (E),
            \end{split}
            \tag{S24}
            \label{S24}
         \end{equation}
         \begin{equation}
            \begin{split}
                w_{A\mathit{2}} (E) & = \frac{1}{2 \pi i} \int_{0}^{2 \pi} d k \partial_k \ln \det \left[ E - A^T (k) \right]
                \\
                & = \frac{1}{2 \pi i} \int_{0}^{2 \pi} d k \partial_k \ln \det \left[ \left(E - A(k) \right)^T\right]
                \\
                & \frac{1}{2 \pi i} \int_{0}^{2 \pi} d k \partial_k \ln \det \left[ E - A(k) \right]
                \\
                & = w(E),
            \end{split}
            \tag{S25}
            \label{S25}
         \end{equation}
         \begin{equation}
            \begin{split}
                w_{A\mathit{3}} (E) & =\frac{1}{2 \pi i} \int_{0}^{2 \pi} d k \partial_k \ln \det \left[ E + A(k) \right]
                \\
                & = \frac{1}{2 \pi i} \int_{0}^{2 \pi} d k \partial_k \ln \{ (-1)^{s} \det \left[ - E - A(k) \right] \}
                \\
                & = \frac{1}{2 \pi i} \int_{0}^{2 \pi} d k \partial_k \ln \det \left[ - E - A(k) \right] ,
            \end{split}
            \tag{S26}
            \label{S26}
         \end{equation}
         \begin{equation}
            \begin{split}
                w_{A\mathit{4}} (E) & = \frac{1}{2 \pi i} \int_{0}^{2 \pi} d k \partial_k \ln \det \left[ E - A^{*} (k) \right]
                \\
                & = \left( \frac{-1}{2 \pi i} \int_{0}^{2 \pi} d k \partial_k \ln \det \left[ E^{*} - A (k) \right]  \right)^{*}
                \\
                & = \frac{-1}{2 \pi i} \int_{0}^{2 \pi} d k \partial_k \ln \det \left[ E^{*} - A (k) \right].
            \end{split}
            \tag{S27}
            \label{S27}
         \end{equation}

      \subsection{TRS}
        According to Eq.(11) in the main text,
          \begin{equation}
             w_{m}^{\mathcal{T}_{+}}(E) = \frac{1}{2 \pi i} \int_{0}^{2 \pi} d k_m \partial_{k_m} \ln \det \left[ U_{\mathcal{T}_{+}} \left(E - H ( e^{\boldsymbol{\mu} + i \mathbf{k}} ) \right) U_{\mathcal{T}_{+}}^{-1} \right].
             \tag{S28}
             \label{S28}
          \end{equation}
        $H ( e^{\boldsymbol{\mu} + i \mathbf{k}} )$ is given in Eq.(18). Thus,
          \begin{equation}
             \begin{split}
                w_{m}^{\mathcal{T}_{+}}(E)  & = \frac{1}{2 \pi i} \int_{0}^{2 \pi} d k_m \partial_{k_m} \ln \det \left[ E - \sum_{\mathbf{j}} U_{\mathcal{T}_{+}} t_{\mathbf{j}} U_{\mathcal{T}_{+}}^{-1} e^{\mathbf{j} \cdot (\boldsymbol{\mu} + i \mathbf{k})} \right].
             \end{split}
             \tag{S29}
             \label{S29}
          \end{equation}
        Substituting Eq.\eqref{S8} into Eq.\eqref{S29}, we get
          \begin{equation}
            \begin{split}
                w_{m}^{\mathcal{T}_{+}} (E)  & = \frac{1}{2 \pi i} \int_{0}^{2 \pi} d k_m \partial_{k_m} \ln \det \left[ E - \sum_{\mathbf{j}}  t_{\mathbf{j}}^{*}  e^{\mathbf{j} \cdot (\boldsymbol{\mu} + i \mathbf{k})} \right]
                \\
                & = \frac{1}{2 \pi i} \int_{0}^{2 \pi} d k_m \partial_{k_m} \ln \det \left[ E - \sum_{\mathbf{j}}  \left( t_{\mathbf{j}}  e^{\mathbf{j} \cdot (\boldsymbol{\mu} - i \mathbf{k})} \right)^{*} \right]
                \\
                & = \frac{-1}{2 \pi i} \int_{0}^{2 \pi} d k_m \partial_{k_m} \ln \det \left[ E^{*} - \sum_{\mathbf{j}}  t_{\mathbf{j}}  e^{\mathbf{j} \cdot (\boldsymbol{\mu} - i \mathbf{k})} \right]
                \\
                & = \frac{1}{2 \pi i} \int_{0}^{2 \pi} d k_m \partial_{k_m} \ln \det \left[ E^{*} - \sum_{\mathbf{j}}  t_{\mathbf{j}}  e^{\mathbf{j} \cdot (\boldsymbol{\mu} + i \mathbf{k})} \right]
                \\
                & = \frac{1}{2 \pi i} \int_{0}^{2 \pi} d k_m \partial_{k_m} \ln \det \left[ E^{*} - H (e^{\boldsymbol{\mu} + i \mathbf{k}}) \right].
            \end{split}
            \tag{S30}
            \label{S30}
          \end{equation}
        Thus, $\tilde{E}_{\mathcal{T}_{+}} = E^{*}$, $\tilde{\boldsymbol{\mu}}_{\mathcal{T}_{+}} = \boldsymbol{\mu}$.

      \subsection{PHS}
          \begin{equation}
            \begin{split}
                w_m^{\mathcal{C}_{-}} (E) & = \frac{1}{2 \pi i} \int_{0}^{2 \pi} d k_m \partial_{k_m} \ln \det \left[ E - \sum_{\mathbf{j}} U_{\mathcal{C}_{-}} t_{\mathbf{j}} U_{\mathcal{C}_{-}}^{-1} e^{\mathbf{j} \cdot (\boldsymbol{\mu} + i \mathbf{k})} \right]
                \\
                & = \frac{1}{2 \pi i} \int_{0}^{2 \pi} d k_m \partial_{k_m} \ln \det \left[ E + \sum_{\mathbf{j}}  t_{-\mathbf{j}}^{T}  e^{\mathbf{j} \cdot (\boldsymbol{\mu} + i \mathbf{k})} \right]
                \\
                & = \frac{1}{2 \pi i} \int_{0}^{2 \pi} d k_m \partial_{k_m} \ln \det \left[ E + \sum_{\mathbf{j}}  t_{\mathbf{j}}  e^{\mathbf{j} \cdot (-\boldsymbol{\mu} - i \mathbf{k})} \right]
                \\
                & = \frac{1}{2 \pi i} \int_{0}^{2 \pi} d k_m \partial_{k_m} \ln \det \left[- E - \sum_{\mathbf{j}}  t_{\mathbf{j}}  e^{\mathbf{j} \cdot (-\boldsymbol{\mu} - i \mathbf{k})} \right]
                \\
                & = \frac{-1}{2 \pi i} \int_{0}^{2 \pi} d k_m \partial_{k_m} \ln \det \left[ -E - \sum_{\mathbf{j}}  t_{\mathbf{j}}  e^{\mathbf{j} \cdot (-\boldsymbol{\mu} + i \mathbf{k})} \right]
                \\
                & = \frac{-1}{2 \pi i} \int_{0}^{2 \pi} d k_m \partial_{k_m} \ln \det \left[- E - H(e^{-\boldsymbol{\mu} + i \mathbf{k}})\right].
            \end{split}
            \tag{S31}
            \label{S31}
          \end{equation}
        Thus, $\tilde{E}_{\mathcal{C}_{-}} = -E$, $\tilde{\boldsymbol{\mu}}_{\mathcal{C}_{-}} = -\boldsymbol{\mu}$.

      \subsection{CS}
        \begin{equation}
            \begin{split}
                w_m^{\mathcal{S}} (E) &  = \frac{1}{2 \pi i} \int_{0}^{2 \pi} d k_m \partial_{k_m} \ln \det \left[ E - \sum_{\mathbf{j}} U_{\mathcal{S}} t_{\mathbf{j}} U_{\mathcal{S}}^{-1} e^{\mathbf{j} \cdot (\boldsymbol{\mu} + i \mathbf{k})} \right]
                \\
                & = \frac{1}{2 \pi i} \int_{0}^{2 \pi} d k_m \partial_{k_m} \ln \det \left[ E + \sum_{\mathbf{j}}  t_{-\mathbf{j}}^{\dagger}  e^{\mathbf{j} \cdot (\boldsymbol{\mu} + i \mathbf{k})} \right]
                \\
                & = \frac{1}{2 \pi i} \int_{0}^{2 \pi} d k_m \partial_{k_m} \ln \det \left[ -E - \sum_{\mathbf{j}}  t_{-\mathbf{j}}^{*}  e^{\mathbf{j} \cdot (\boldsymbol{\mu} + i \mathbf{k})} \right]
                \\
                & = \frac{1}{2 \pi i} \int_{0}^{2 \pi} d k_m \partial_{k_m} \ln \det \left[ -E - \sum_{\mathbf{j}}  \left(t_{\mathbf{j}}  e^{\mathbf{j} \cdot (-\boldsymbol{\mu} + i \mathbf{k})}\right)^{*} \right]
                \\
                & = \frac{-1}{2 \pi i} \int_{0}^{2 \pi} d k_m \partial_{k_m} \ln \det \left[ -E^{*} - \sum_{\mathbf{j}}  t_{\mathbf{j}}  e^{\mathbf{j} \cdot (-\boldsymbol{\mu} + i \mathbf{k})} \right]
                \\
                & = \frac{-1}{2 \pi i} \int_{0}^{2 \pi} d k_m \partial_{k_m} \ln \det \left[ -E^{*} - H(e^{-\boldsymbol{\mu} + i \mathbf{k}}) \right].
            \end{split}
            \tag{S32}
            \label{S32}
        \end{equation}
        Thus, $\tilde{E}_{\mathcal{S}} = -E^{*}$, $\tilde{\boldsymbol{\mu}}_{\mathcal{S}} = -\boldsymbol{\mu}$.

      \subsection{TRS$^\dagger$}
          \begin{equation}
             \begin{split}
                w_m^{\mathcal{C}_{+}} (E) & = \frac{1}{2 \pi i} \int_{0}^{2 \pi} d k_m \partial_{k_m} \ln \det \left[ E - \sum_{\mathbf{j}} U_{\mathcal{C}_{+}} t_{\mathbf{j}} U_{\mathcal{C}_{+}}^{-1} e^{\mathbf{j} \cdot (\boldsymbol{\mu} + i \mathbf{k})} \right]
                \\
                & = \frac{1}{2 \pi i} \int_{0}^{2 \pi} d k_m \partial_{k_m} \ln \det \left[ E - \sum_{\mathbf{j}}  t_{-\mathbf{j}}^{T}  e^{\mathbf{j} \cdot (\boldsymbol{\mu} + i \mathbf{k})} \right]
                \\
                & = \frac{1}{2 \pi i} \int_{0}^{2 \pi} d k_m \partial_{k_m} \ln \det \left[ E - \sum_{\mathbf{j}}  t_{\mathbf{j}}  e^{\mathbf{j} \cdot (-\boldsymbol{\mu} - i \mathbf{k})} \right]
                \\
                & = \frac{-1}{2 \pi i} \int_{0}^{2 \pi} d k_m \partial_{k_m} \ln \det \left[ E - \sum_{\mathbf{j}}  t_{\mathbf{j}}  e^{\mathbf{j} \cdot (-\boldsymbol{\mu} + i \mathbf{k})} \right]
                \\
                & = \frac{-1}{2 \pi i} \int_{0}^{2 \pi} d k_m \partial_{k_m} \ln \det \left[ E - H(e^{-\boldsymbol{\mu} + i \mathbf{k}}) \right]
             \end{split}
             \tag{S33}
             \label{S33}
          \end{equation}
          Thus, $\tilde{E}_{\mathcal{C}_{+}} = E$, $\tilde{\boldsymbol{\mu}}_{\mathcal{C}_{+}} = -\boldsymbol{\mu}$.

       \subsection{PHS$^\dagger$}
          \begin{equation}
             \begin{split}
                w_{m}^{\mathcal{T}_{-}}(E)  & = \frac{1}{2 \pi i} \int_{0}^{2 \pi} d k_m \partial_{k_m} \ln \det \left[ E - \sum_{\mathbf{j}} U_{\mathcal{T}_{-}} t_{\mathbf{j}} U_{\mathcal{T}_{-}}^{-1} e^{\mathbf{j} \cdot (\boldsymbol{\mu} + i \mathbf{k})} \right]
                \\
                & = \frac{1}{2 \pi i} \int_{0}^{2 \pi} d k_m \partial_{k_m} \ln \det \left[ E + \sum_{\mathbf{j}}  t_{\mathbf{j}}^{*}  e^{\mathbf{j} \cdot (\boldsymbol{\mu} + i \mathbf{k})} \right]
                \\
                & = \frac{1}{2 \pi i} \int_{0}^{2 \pi} d k_m \partial_{k_m} \ln \det \left[ -E - \sum_{\mathbf{j}}  t_{\mathbf{j}}^{*}  e^{\mathbf{j} \cdot (\boldsymbol{\mu} + i \mathbf{k})} \right]
                \\
                & = \frac{-1}{2 \pi i} \int_{0}^{2 \pi} d k_m \partial_{k_m} \ln \det \left[ -E^{*} - \sum_{\mathbf{j}}  t_{\mathbf{j}}  e^{\mathbf{j} \cdot (\boldsymbol{\mu} - i \mathbf{k})} \right]
                \\
                & = \frac{1}{2 \pi i} \int_{0}^{2 \pi} d k_m \partial_{k_m} \ln \det \left[ -E^{*} - \sum_{\mathbf{j}}  t_{\mathbf{j}}  e^{\mathbf{j} \cdot (\boldsymbol{\mu} + i \mathbf{k})} \right] 
                \\
                & = \frac{1}{2 \pi i} \int_{0}^{2 \pi} d k_m \partial_{k_m} \ln \det \left[ -E^{*} - H(e^{\boldsymbol{\mu} + i \mathbf{k}}) \right].
            \end{split}
             \tag{S34}
             \label{S34}
          \end{equation}
          Thus, $\tilde{E}_{\mathcal{T}_{-}} = -E^{*}$, $\tilde{\boldsymbol{\mu}}_{\mathcal{T}_{-}} = \boldsymbol{\mu}$.

      \subsection{SLS}
          \begin{equation}
            \begin{split}
                w_m^{\Gamma} (E) &  = \frac{1}{2 \pi i} \int_{0}^{2 \pi} d k_m \partial_{k_m} \ln \det \left[ E - \sum_{\mathbf{j}} U_{\Gamma} t_{\mathbf{j}} U_{\Gamma}^{-1} e^{\mathbf{j} \cdot (\boldsymbol{\mu} + i \mathbf{k})} \right]
                \\
                & = \frac{1}{2 \pi i} \int_{0}^{2 \pi} d k_m \partial_{k_m} \ln \det \left[ E + \sum_{\mathbf{j}}  t_{\mathbf{j}}  e^{\mathbf{j} \cdot (\boldsymbol{\mu} + i \mathbf{k})} \right]
                \\
                & = \frac{1}{2 \pi i} \int_{0}^{2 \pi} d k_m \partial_{k_m} \ln \det \left[ - E - \sum_{\mathbf{j}}  t_{\mathbf{j}}  e^{\mathbf{j} \cdot (\boldsymbol{\mu} + i \mathbf{k})} \right]
                \\
                & = \frac{1}{2 \pi i} \int_{0}^{2 \pi} d k_m \partial_{k_m} \ln \det \left[ -E - H(e^{\boldsymbol{\mu} + i \mathbf{k}}) \right].
            \end{split}
            \tag{S35}
            \label{S35}
          \end{equation}
          Thus, $\tilde{E}_{\Gamma} = -E$, $\tilde{\boldsymbol{\mu}}_{\Gamma} = \boldsymbol{\mu}$.

      \subsection{Pseudo Hermitian Symmetry}
          \begin{equation}
            \begin{split}
                w_m^{\eta} (E) &  = \frac{1}{2 \pi i} \int_{0}^{2 \pi} d k_m \partial_{k_m} \ln \det \left[ E - \sum_{\mathbf{j}} \eta t_{\mathbf{j}} \eta^{-1} e^{\mathbf{j} \cdot (\boldsymbol{\mu} + i \mathbf{k})} \right]
                \\
                & = \frac{1}{2 \pi i} \int_{0}^{2 \pi} d k_m \partial_{k_m} \ln \det \left[ E - \sum_{\mathbf{j}}  t_{-\mathbf{j}}^{\dagger}  e^{\mathbf{j} \cdot (\boldsymbol{\mu} + i \mathbf{k})} \right]
                \\
                & = \frac{1}{2 \pi i} \int_{0}^{2 \pi} d k_m \partial_{k_m} \ln \det \left[ E - \sum_{\mathbf{j}}  t_{\mathbf{j}}^{*}  e^{\mathbf{j} \cdot (-\boldsymbol{\mu} - i \mathbf{k})} \right]
                \\
                & = \frac{-1}{2 \pi i} \int_{0}^{2 \pi} d k_m \partial_{k_m} \ln \det \left[ E^{*} - \sum_{\mathbf{j}}  t_{\mathbf{j}}  e^{\mathbf{j} \cdot (-\boldsymbol{\mu} + i \mathbf{k})} \right]
                \\
                & = \frac{-1}{2 \pi i} \int_{0}^{2 \pi} d k_m \partial_{k_m} \ln \det \left[ E^{*} - H(e^{-\boldsymbol{\mu} + i \mathbf{k}}) \right].
            \end{split}
            \tag{S36}
            \label{S36}
          \end{equation}
          Thus, $\tilde{E}_{\eta} = E^{*}$, $\tilde{\boldsymbol{\mu}}_{\eta} = -\boldsymbol{\mu}$.
          \par
          \vspace{15pt}

          Now, we get the results in table$\mathbf{I}$ in main text.

   \section{Examples}

      \subsection{PHS}
\begin{figure}
            \centering
            \subfigure[]{\includegraphics[scale=0.45]{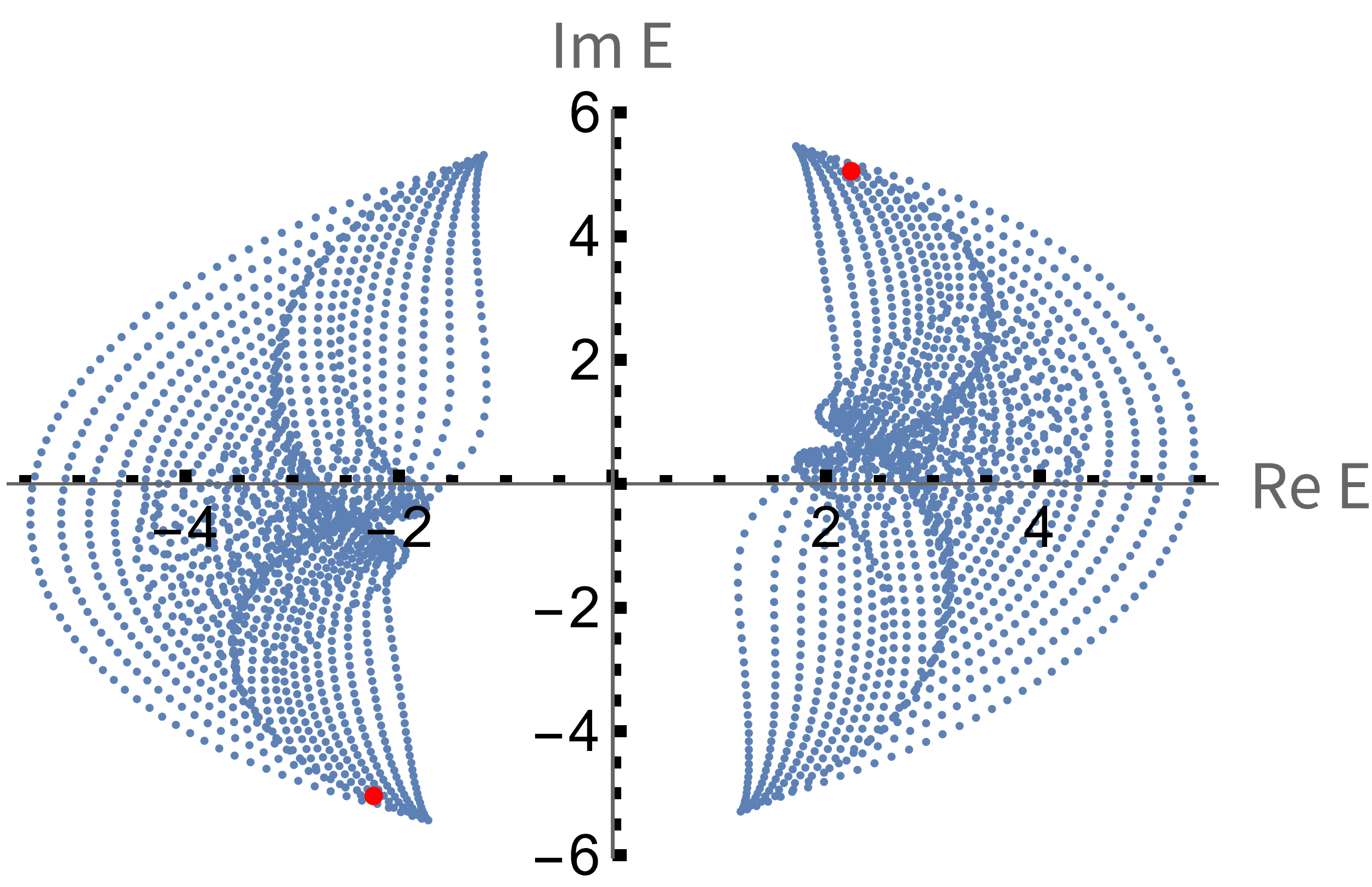} \label{figs1a}}
            \quad
            \subfigure[]{\includegraphics[scale=0.3]{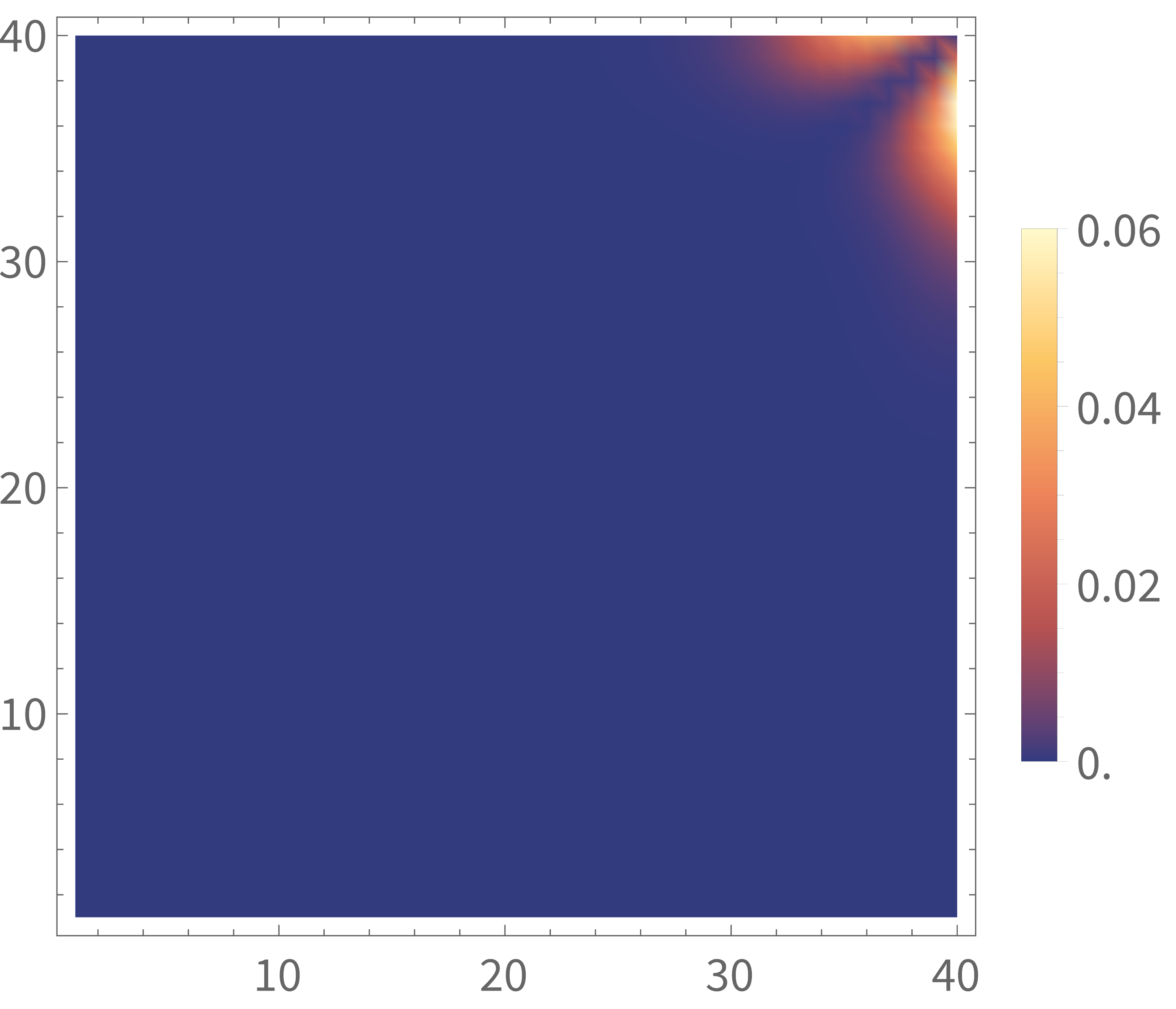} \label{figs1b}}
            \quad
            \subfigure[]{\includegraphics[scale=0.3]{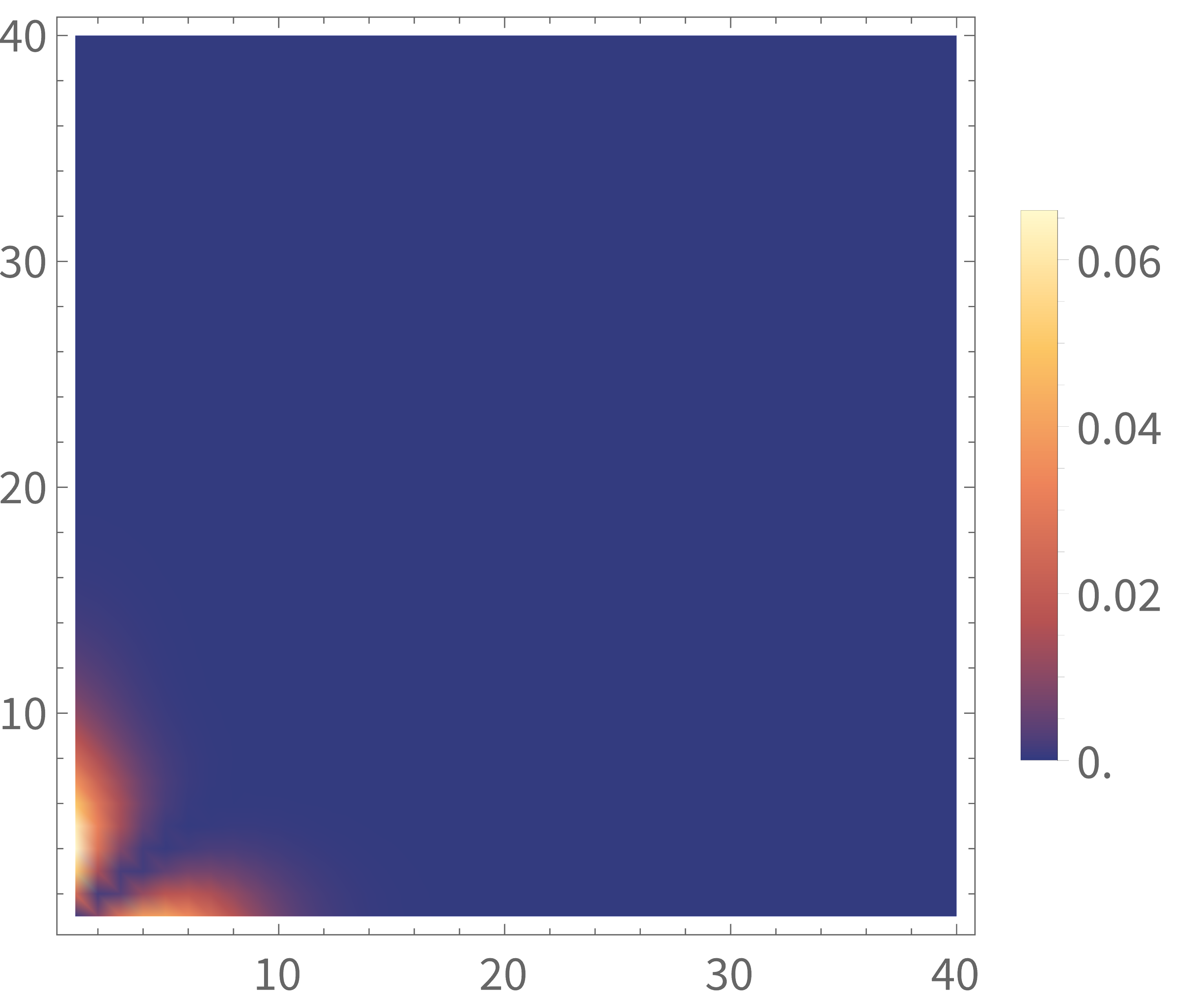} \label{figs1c}}
            \caption{(a)The blue dots are the spectrum of the system given by Eq.\eqref{S37} under OBC. Two red dots are $E_{PHS1} = 2.24 + 5.05 i$ and $E_{PHS2} = -2.24 - 5.05 i$ respectively. (b)The distribution of wave function corresponding to $E_{PHS1}$. (c)The distribution of wave function corresponding to $E_{PHS2}$. \  Values of parameters are $m=1.5$, $\gamma=1$, $t_x=3$, $t_y=2.5$, $t=1$, $\tilde{t}=2$. The size of the system is $40\times40$.}
           
            \label{figs1}
          \end{figure}

        We consider a $2$-dimensional non-Hermitian system with PHS, the Hamiltonian under PBC is
          \begin{equation}
             H_{PHS} (k_x,k_y) = 
               \begin{pmatrix}
                  m + i \gamma + t_x e^{i k_x} + t_y e^{i k_y}  &  t  \\
                  \tilde{t}  & -m - i \gamma - t_x e^{i k_x} - t_y e^{i k_y}
               \end{pmatrix}.
               \tag{S37}
               \label{S37}
          \end{equation}
        The operator of PHS is 
          \begin{equation}
            \hat{\mathcal{C}}_{-} = U_{\mathcal{C}_{-}} =
              \begin{pmatrix}
                 0 & 1 \\
                 -1 & 0
              \end{pmatrix}.
              \tag{S38}
              \label{S38}
          \end{equation}
        The spectrum under OBC for specific values of parameters are given in Fig.\ref{figs1}. For two eigenenergies, $E_{PHS1} = 2.24 + 5.05 i$ and $E_{PHS2} = -2.24 - 5.05 i$, belong to the OBC spectrum, $E_{PHS1} = - E_{PHS2}$ and eigenstates corresponding to them are localized at the opposite boundary.

      \subsection{CS}
\begin{figure}
            \centering
            \subfigure[]{\includegraphics[scale=0.45]{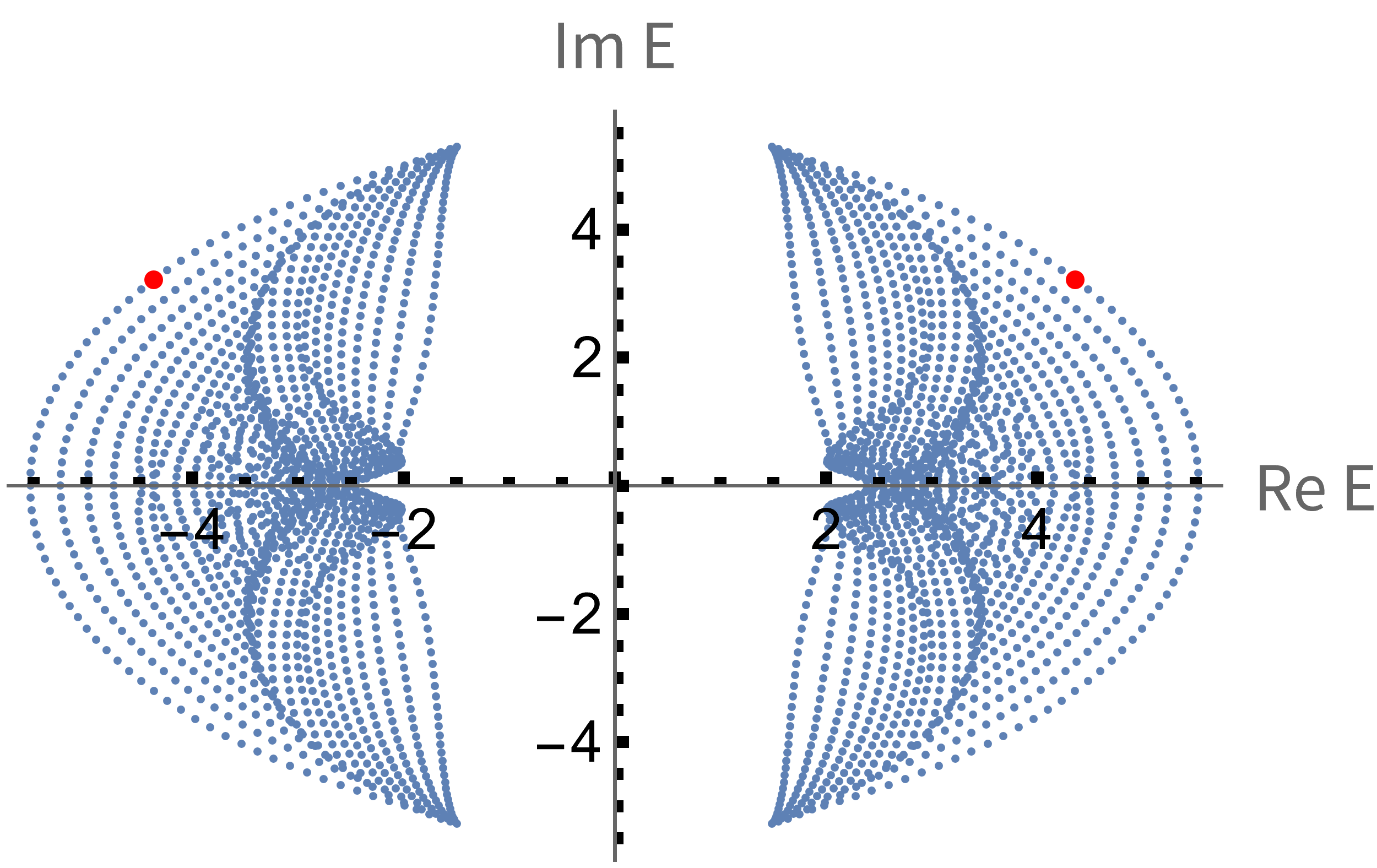} \label{figs2a}}
            \quad
            \subfigure[]{\includegraphics[scale=0.3]{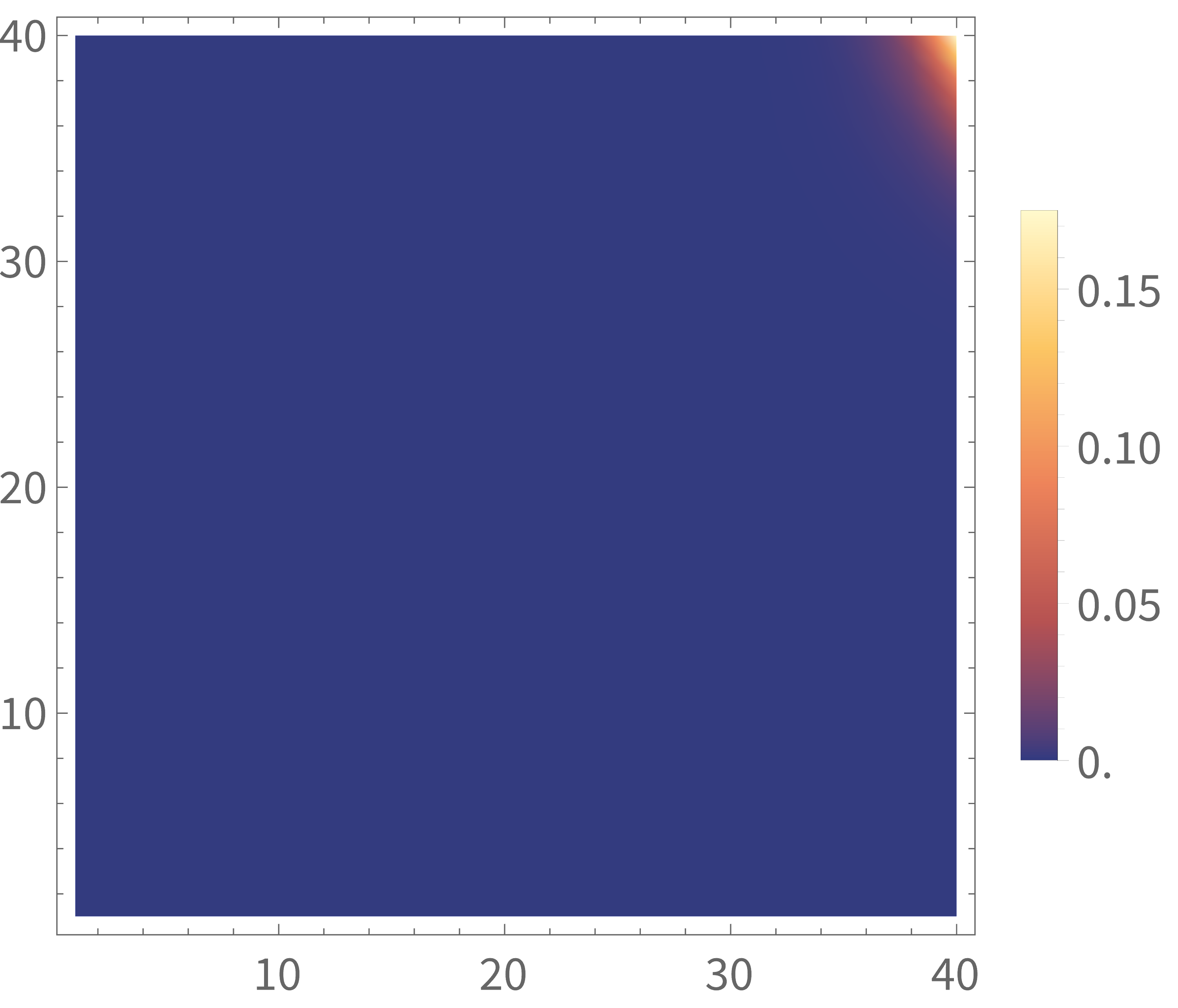} \label{figs2b}}
            \quad
            \subfigure[]{\includegraphics[scale=0.3]{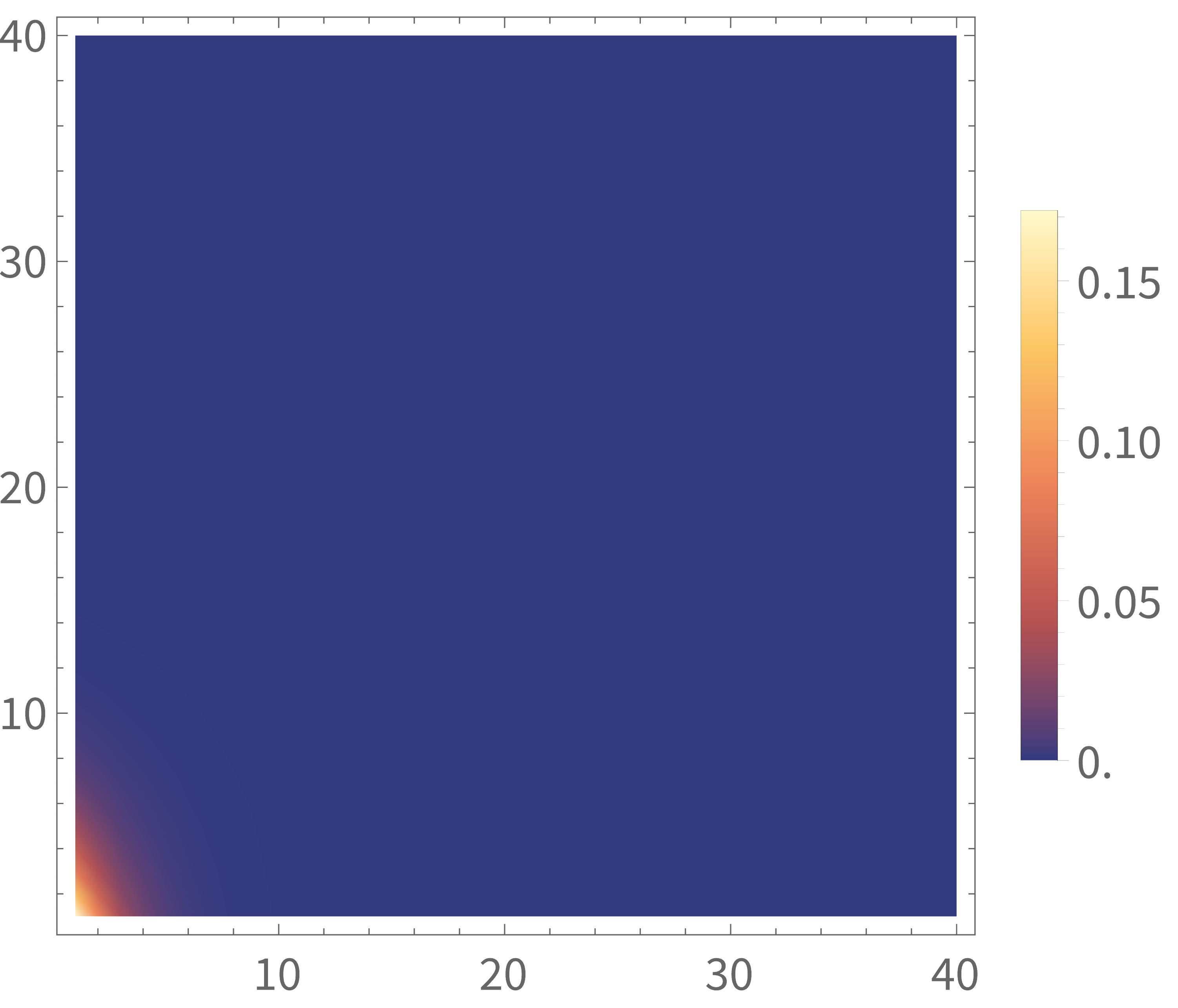} \label{figs2c}}
            \caption{(a)The blue dots are the spectrum of the system given by Eq.\eqref{S39} under OBC. Two red dots are $E_{CS1} = 4.36 + 3.21 i$ and $E_{CS2} = -4.36 + 3.21 i$ respectively. (b)The distribution of wave function corresponding to $E_{CS1}$. (c)The distribution of wave function corresponding to $E_{CS2}$. \  Values of parameters are $m=1.5$, $t_x=3$, $t_y=2.5$, $t=1$, $\tilde{t}=2$. The size of the system is $40\times40$.}
            \label{figs2}
          \end{figure}

        Setting $\gamma = 0$ in Eq.\eqref{S37}, the Hamiltonian given in Eq.\eqref{S37} has TRS. Thus, we take the Hamiltonian with CS under PBC as 
          \begin{equation}
            H_{CS} (k_x,k_y) = 
            \begin{pmatrix}
               m + t_x e^{i k_x} + t_y e^{i k_y}  &  t  \\
               \tilde{t}  & -m - t_x e^{i k_x} - t_y e^{i k_y}
            \end{pmatrix},
            \tag{S39}
            \label{S39}
          \end{equation} 
        and the operator of CS is 
          \begin{equation}
             \hat{\mathcal{S}} = U_{\mathcal{S}} K =
               \begin{pmatrix}
                 0  &  1 \\
                 -1 &  0
               \end{pmatrix}
               K.
               \tag{S40}
               \label{S40}
          \end{equation}
        We still take values of the parameters as $m=1.5$, $t_x=3$, $t_y=2.5$, $t=1$, $\tilde{t}=2$. In this case, for two eigenenergies, $E_{CS1} = 4.36 + 3.21 i$ and $E_{CS2} = - E_{CS1}^{*} = -4.36 + 3.21 i$, eigenstates corresponding to $E_{CS1}$ and $E_{CS2}$ are localizing at the opposite boundary~(See Fig.\ref{figs2}).

      \subsection{PHS$^\dagger$}
\begin{figure}
            \centering
            \subfigure[]{\includegraphics[scale=0.45]{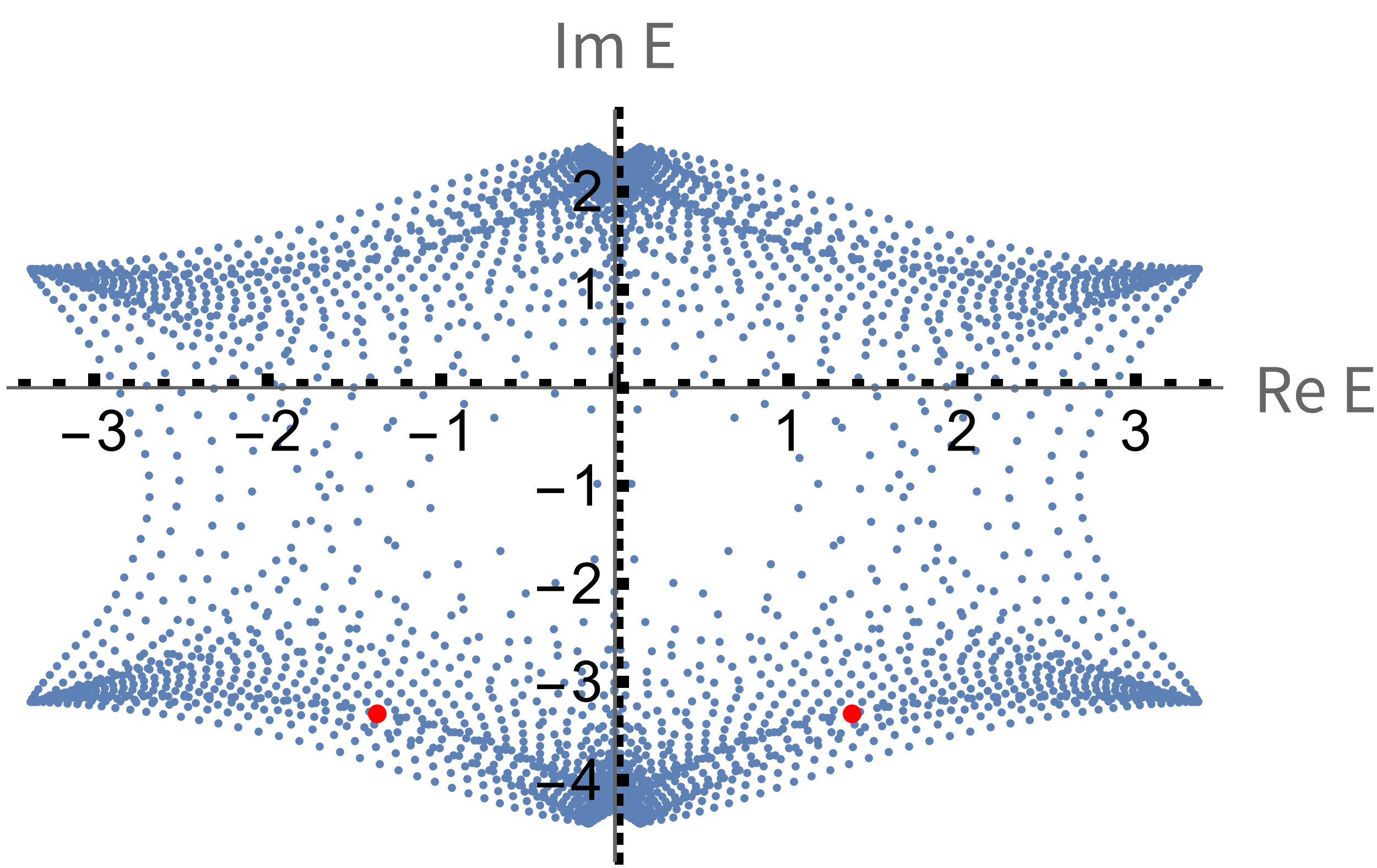} \label{figs3a}}
            \quad
            \subfigure[]{\includegraphics[scale=0.3]{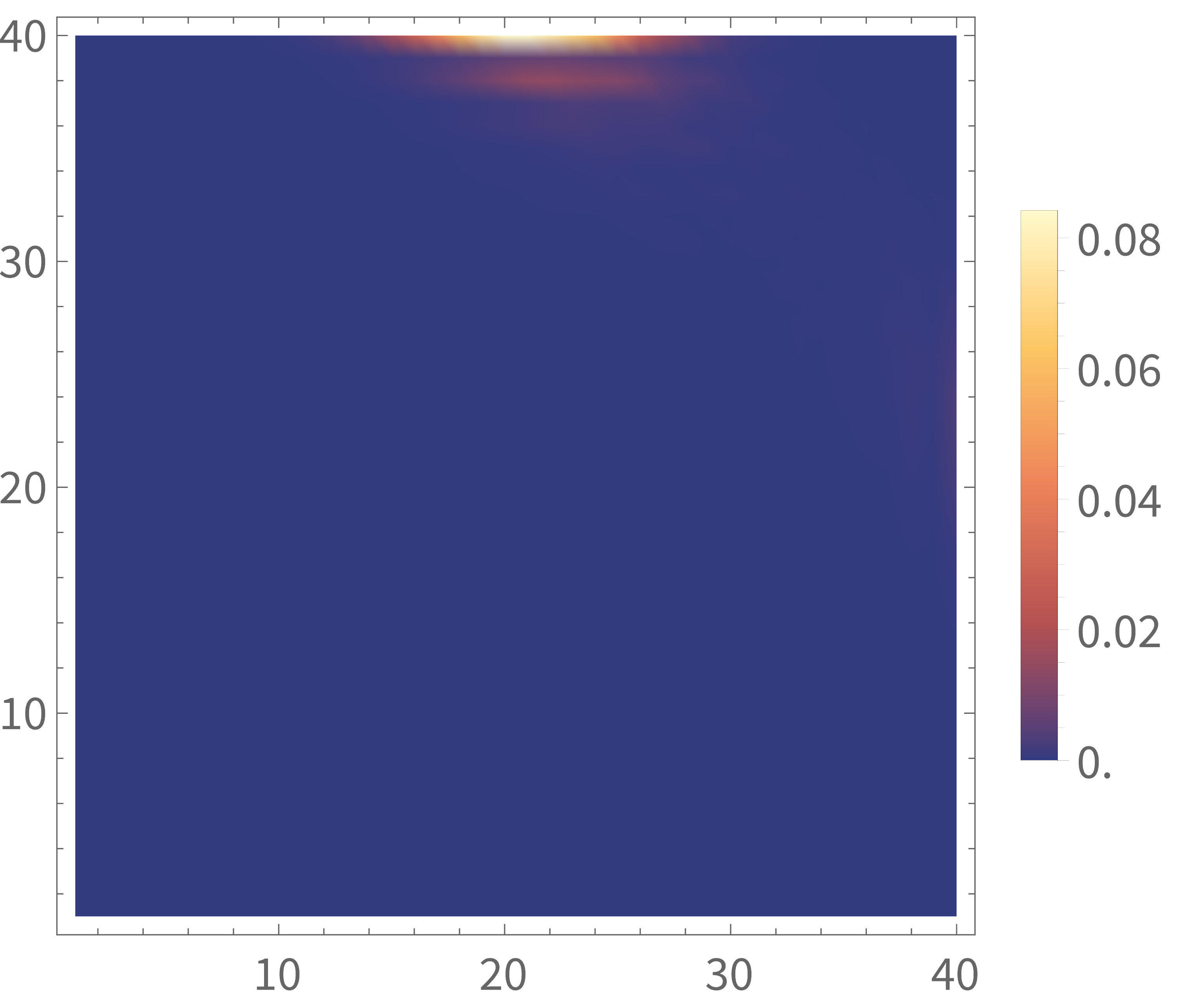} \label{figs3b}}
            \quad
            \subfigure[]{\includegraphics[scale=0.3]{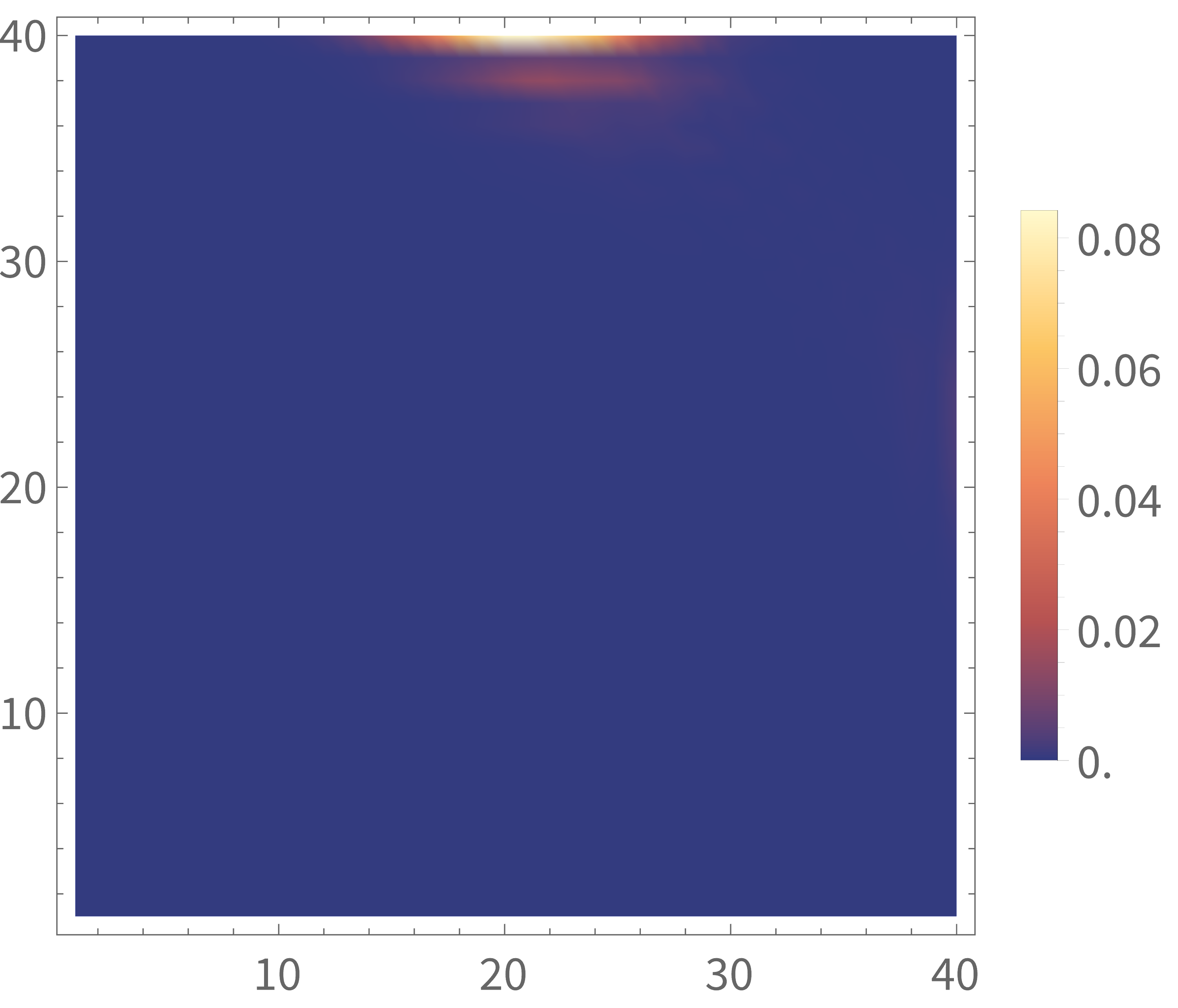} \label{figs3c}}
            \caption{(a)The blue dots are the spectrum of the system given by Eq.\eqref{S41} under OBC. Two red dots are $E_{PHS^{\dagger}1} = 1.37 - 3.33 i$ and $E_{PHS^{\dagger}2} = -1.37 - 3.33 i $ respectively. (b)The distribution of wave function corresponding to $E_{PHS^{\dagger}1}$. (c)The distribution of wave function corresponding to $E_{PHS^{\dagger}2}$. \  Values of parameters are $\tilde{t}_0 = -1$, $t_0 = 3$, $t_x = 2$, $t_y = 2.3$, $\gamma = 2$. The size of the system is $40\times40$.}
            \label{figs3}
          \end{figure}

        Consider the $2$-dimensional non-Hermitian Hamiltonian under PBC,
          \begin{equation}
            H_{PHS^{\dagger}} (k_x,k_y) = 
              \begin{pmatrix}
                 i \gamma          &  t_0 + t_x e^{-i k_x} + t_y e^{-i k_y}  \\
                 \tilde{t}_0 + t_x e^{-i k_x} + t_y e^{-i k_y}  &  -2 i \gamma
              \end{pmatrix}.
              \tag{S41}
              \label{S41}
          \end{equation}
        This system has PHS$^\dagger$ and the operator of PHS$^\dagger$ is
          \begin{equation}
            \hat{\mathcal{T}}_{-} = U_{\mathcal{T}_{-}} = 
              \begin{pmatrix}
                1  &   0  \\
                0  &  -1
              \end{pmatrix}.
              \tag{S42}
              \label{S42}
          \end{equation}
        The spectrum under OBC is given in Fig.\ref{figs3}. Eigenstate corresponding to $E_{PHS^{\dagger}1} = 1.37 - 3.33 i$ is localized at the same boundary with the eigenstate corresponding to $E_{PHS^{\dagger}2} = - E_{PHS^{\dagger}1}^{*} = -1.37 - 3.33 i $.

      \subsection{SLS}
\begin{figure}
              \centering
              \subfigure[]{\includegraphics[scale=0.45]{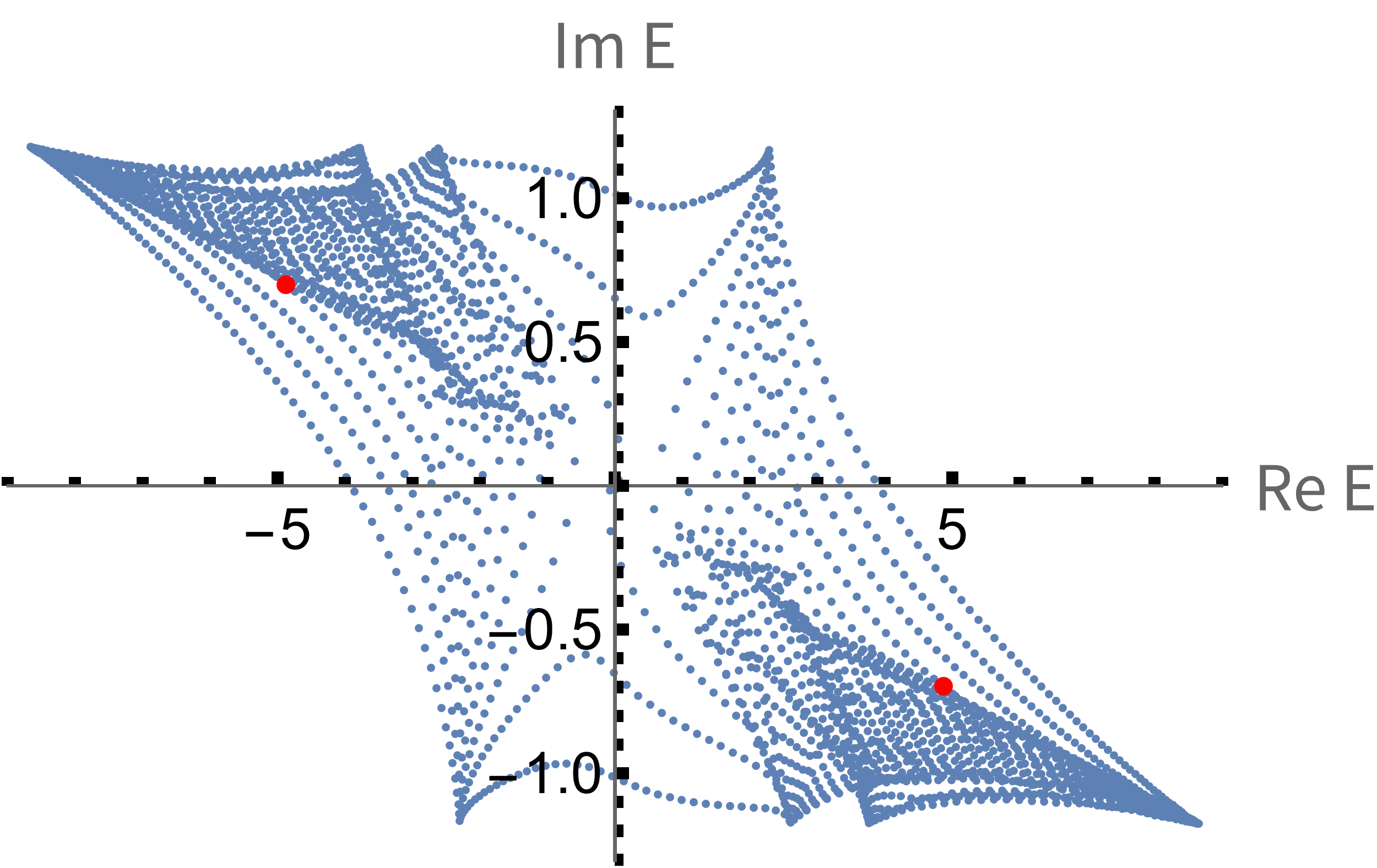} \label{fig4a}}
              \quad
              \subfigure[]{\includegraphics[scale=0.3]{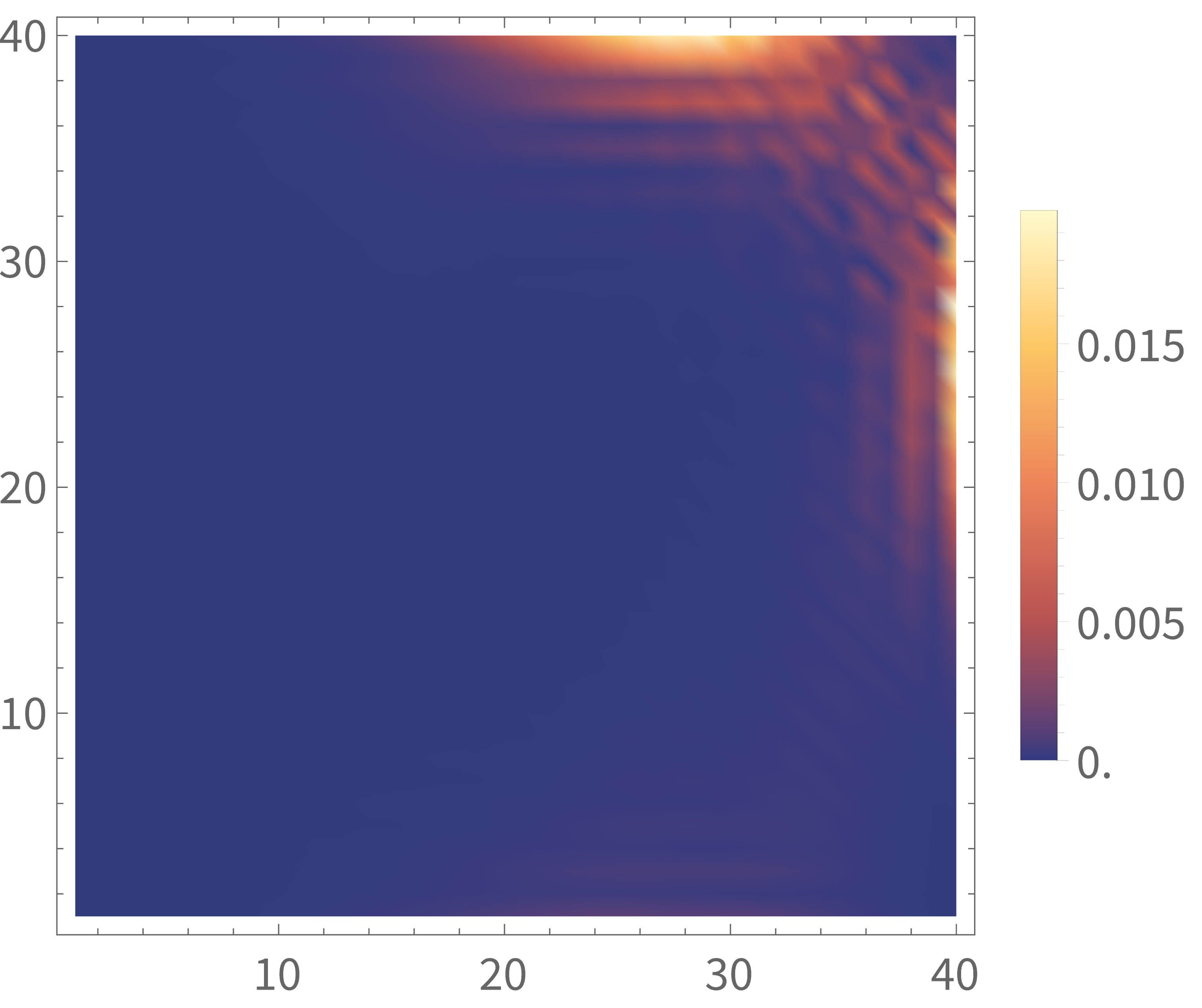} \label{fig4b}}
              \quad
              \subfigure[]{\includegraphics[scale=0.3]{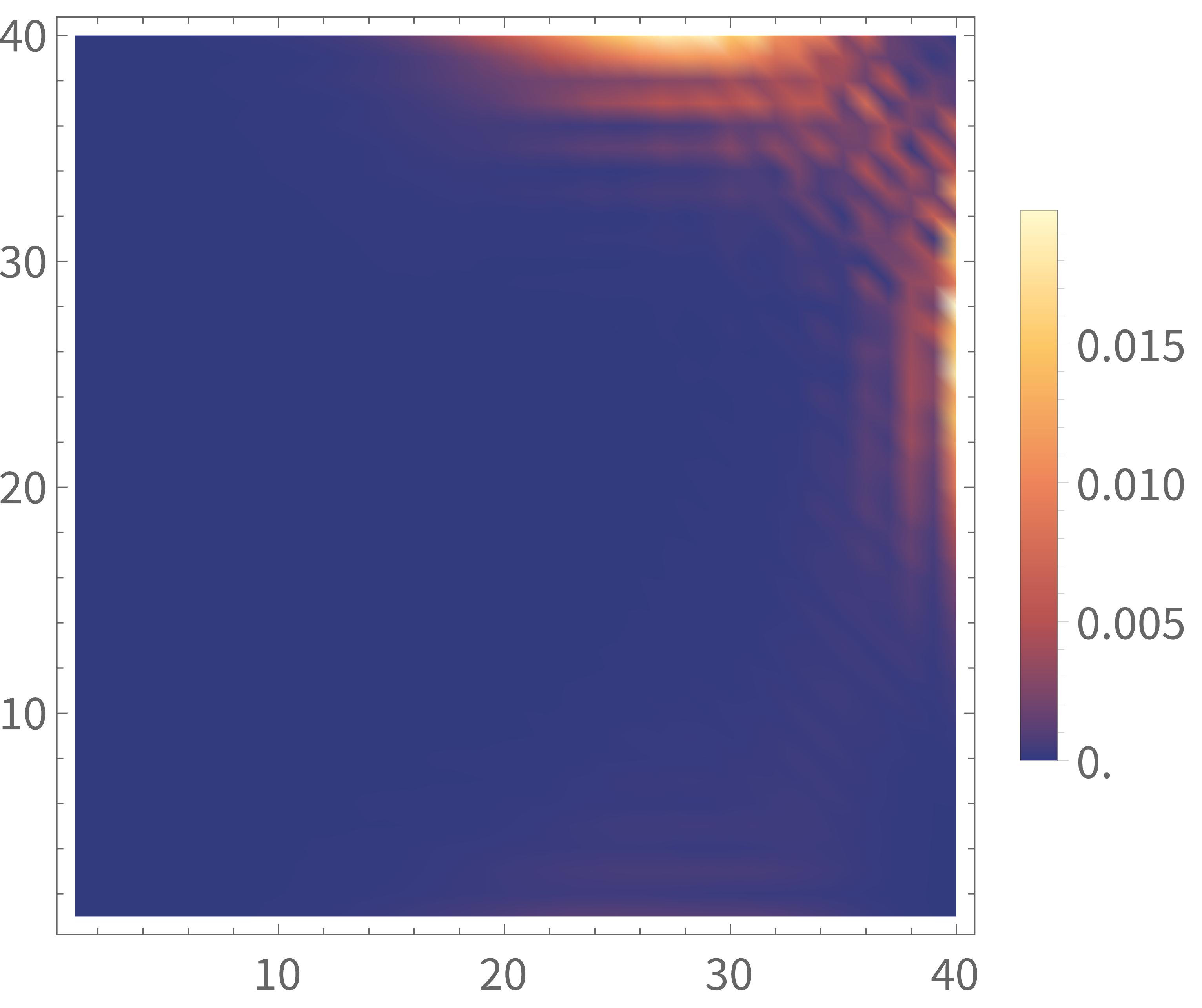} \label{fig4c}}
              \caption{(a)The blue dots are the spectrum of the system given by Eq.\eqref{S43} under OBC. Two red dots are $E_{\Gamma 1} = -4.87 + 0.70 i$ and $E_{\Gamma 2} = 4.87 - 0.70 i $ respectively. (b)The distribution of wave function corresponding to $E_{\Gamma 1}$. (c)The distribution of wave function corresponding to $E_{\Gamma 2}$. \  Values of parameters are $t_0 = 3$, $m=1.5$, $\gamma_1 = 1$, $\gamma_2 = -2$, $t_1 = 2$, $t_{-1} = 3$, $w_1 = 2.3$, $w_{-1} = 4$. The size of the system is $40\times40$.}
              \label{figs4}
            \end{figure}

         We consider the non-Hermitian system with SLS, whose Hamiltonian under PBC is
           \begin{equation}
              H_{\Gamma} (k_x,k_y) = 
                \begin{pmatrix}
                    0   &   t_0 + m + i \gamma_1 + t_{-1} e^{-i k_x} + w_{-1} e^{-i k_y} \\
                    t_0 - m + i \gamma_2 + t_{1} e^{i k_x} + w_{1} e^{i k_y}  & 0 
                \end{pmatrix},
                \tag{S43}
                \label{S43}
           \end{equation}
          and the operator of SLS is 
            \begin{equation}
               \hat{\Gamma} = U_{\Gamma} K = 
                  \begin{pmatrix}
                      1  &  0  \\
                      0  & -1 
                  \end{pmatrix}
                  K.
                  \tag{S44}
                  \label{S44}
            \end{equation}
          The OBC spectrum under specific values of parameters are shown in Fig.\ref{figs4}. Two eigenstates corresponding to $E_{\Gamma 1} = -4.87 + 0.70 i$ and $E_{\Gamma 2}= -E_{\Gamma 1} = 4.87 - 0.70 i$ respectively are localized at the same boundary. 

        \subsection{Pseudo Hermitian Symmetry}
\begin{figure}
              \centering
              \subfigure[]{\includegraphics[scale=0.45]{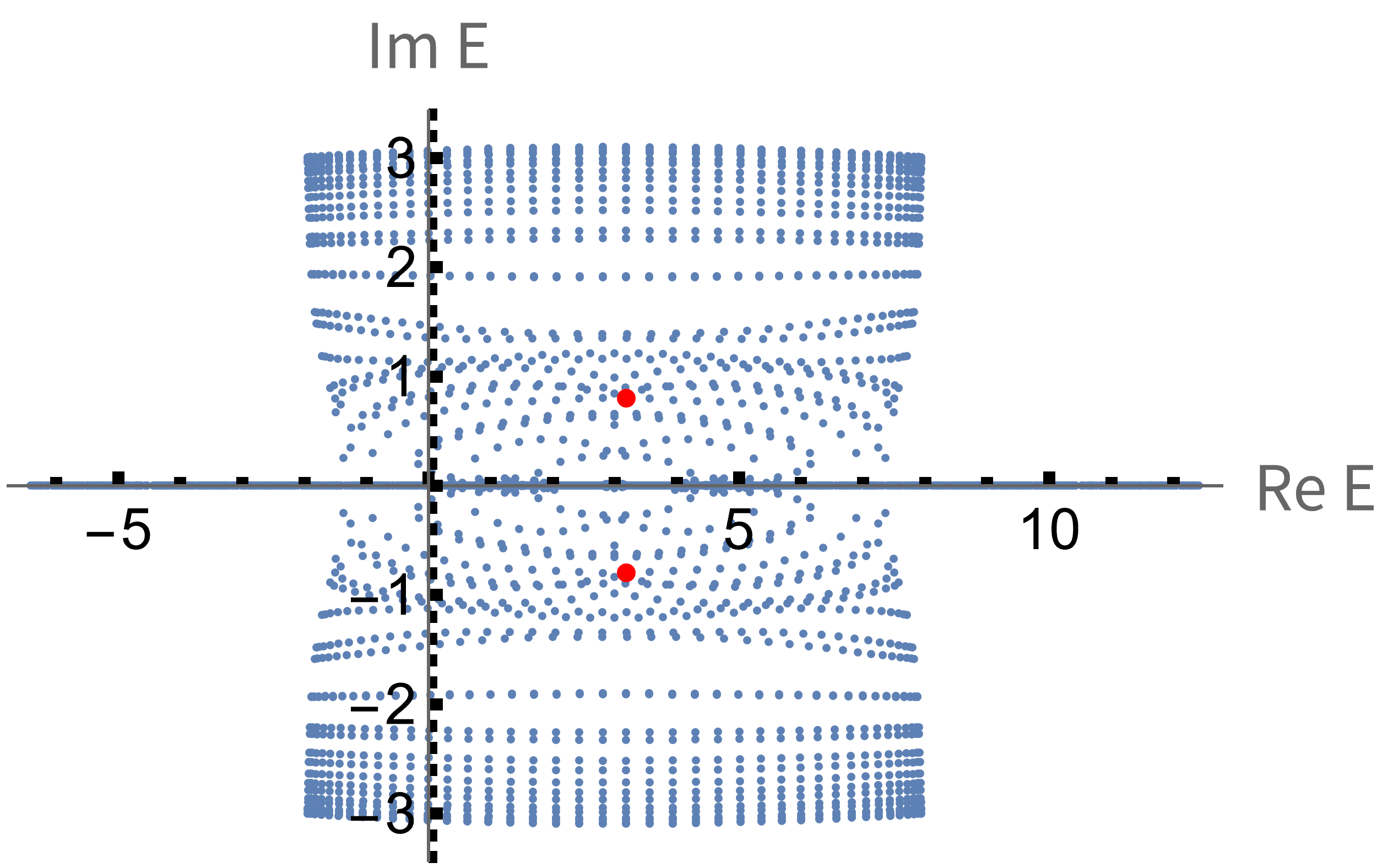} \label{fig5a}}
              \quad
              \subfigure[]{\includegraphics[scale=0.3]{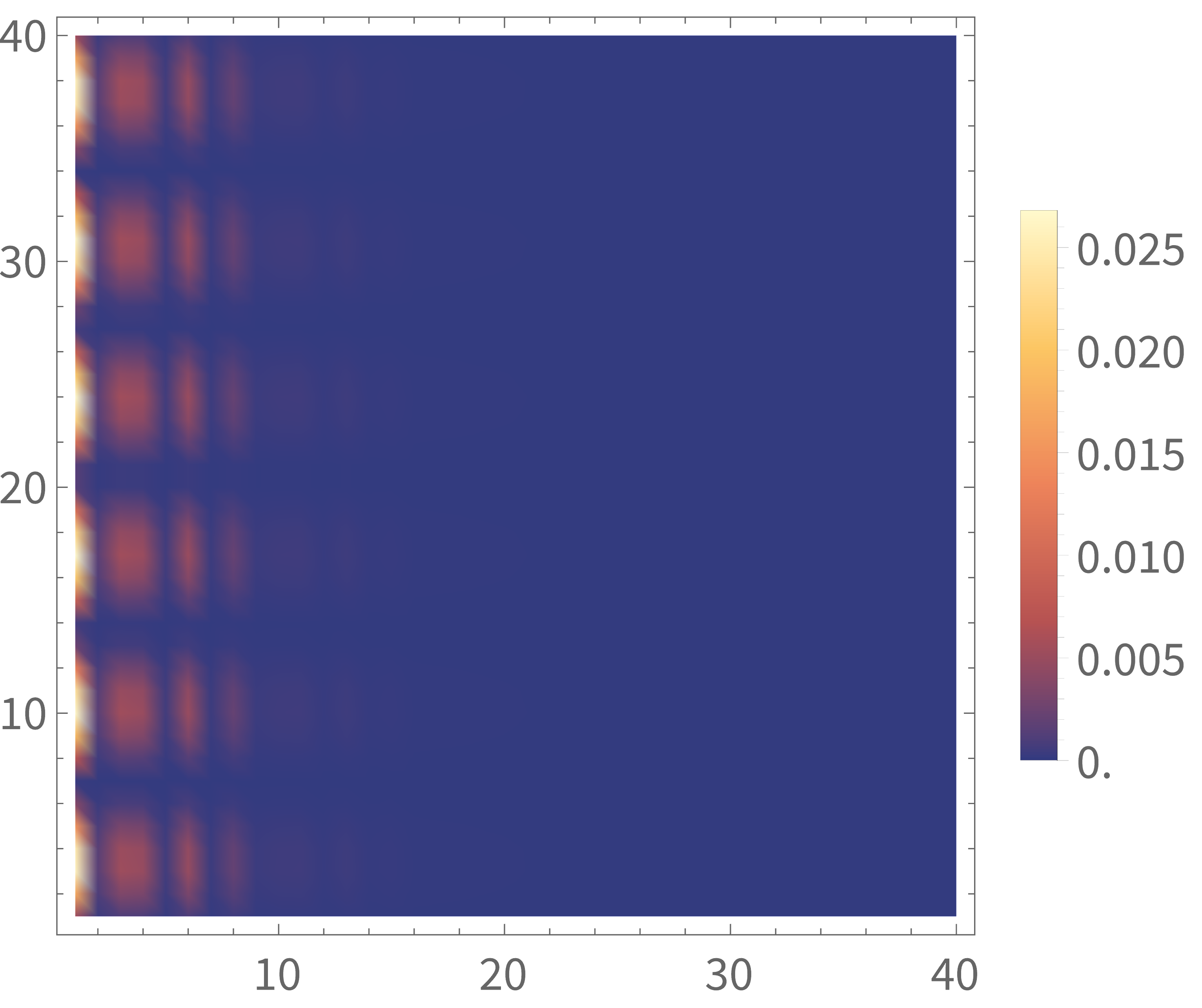} \label{fig5b}}
              \quad
              \subfigure[]{\includegraphics[scale=0.3]{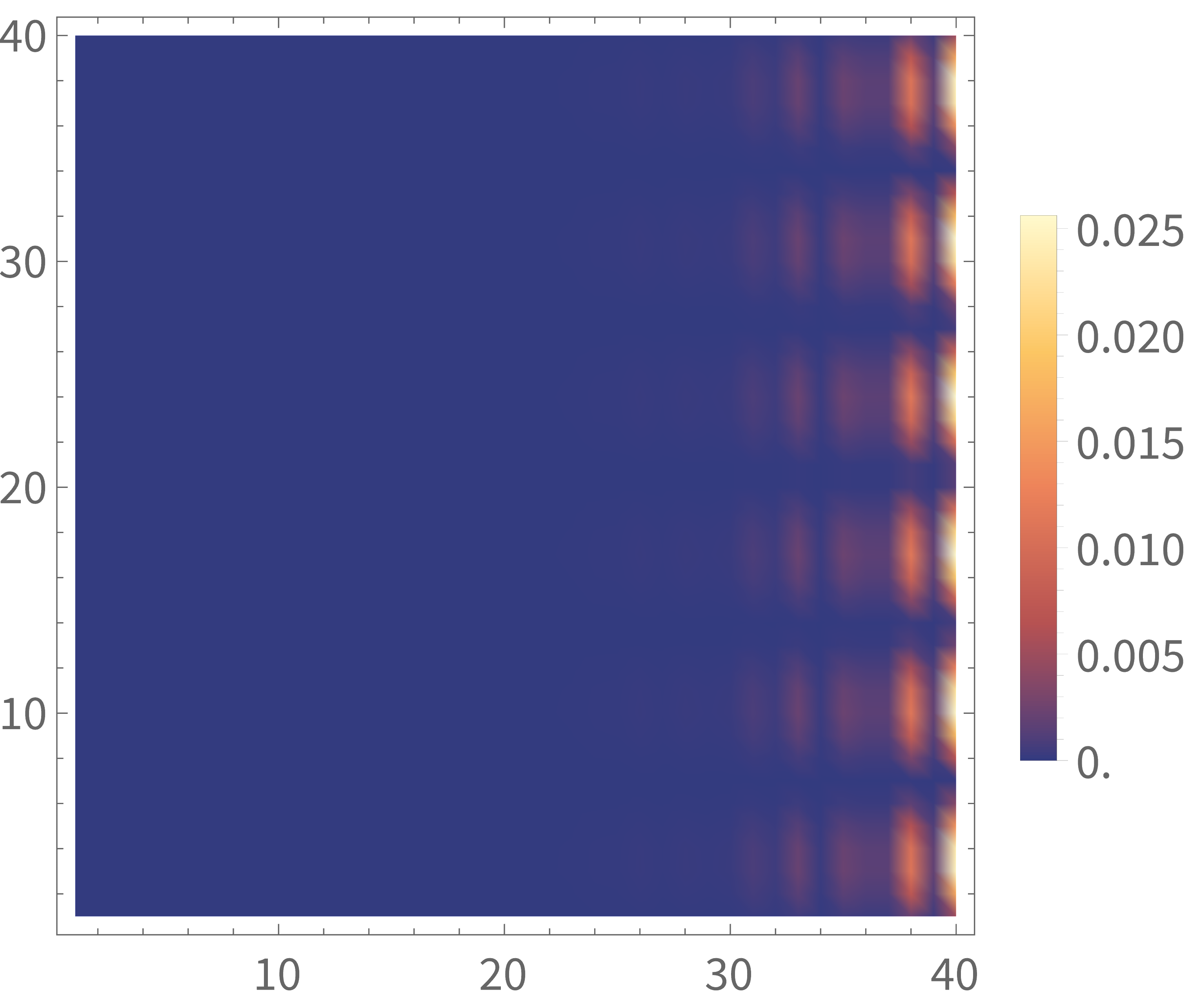} \label{fig5c}}
              \caption{(a)The blue dots are the spectrum of the system given by Eq.\eqref{S45} under OBC. Two red dots are $E_{\eta 1} = 3.19 + 0.80 i$ and $E_{\eta 2} = 3.19 - 0.80 i $ respectively. (b)The distribution of wave function corresponding to $E_{\eta 1}$. (c)The distribution of wave function corresponding to $E_{\eta 2}$. \  Values of parameters are $t_0 = 3$, $\gamma = 2$, $m_1 = 1.5$, $m_2 = -3$, $t_1 = 2$, $t_{-1} = 3$, $w=2.3$. The size of the system is $40\times40$.}
              \label{figs5}
            \end{figure}

          Consider a non-Hermitian system with pseudo Hermitian symmetry, whose Hamiltonian under PBC is
            \begin{equation}
               H_{\eta} (k_x,k_y) = 
                 \begin{pmatrix}
                   t_0 + i \gamma + t_{-1} e^{-i k_x} + t_{1} e^{i k_x}  &
                   m_1 + w e^{-i k_y} + w e^{i k_y}
                   \\
                   m_2 + w e^{-i k_y} + w e^{i k_y} &
                   t_0 - i \gamma + t_{-1} e^{i k_x} + t_{1} e^{-i k_x}  
                 \end{pmatrix},
                 \tag{S45}
                 \label{S45}
            \end{equation}
          and the operator of pseudo Hermitian symmetry is
            \begin{equation}
              \hat{\eta} = U_{\eta} = 
                \begin{pmatrix}
                   0  &  1  \\
                   1  &  0
                \end{pmatrix}.
                \tag{S46}
                \label{S46}
            \end{equation}
          The OBC specific of this system under specific values of parameters is shown in Fig.\ref{figs5}. In this system, the eigenstate corresponding to $E_{\eta 1} = 3.19 + 0.80 i$ and the eigenstate corresponding to $E_{\eta 2} = E_{\eta 1}^{*} = 3.19 - 0.80 i$ are localized at the opposite boundary.

    \section{Examples of  Bidirectional Skin Effect in a $2$-dimensional non-Hermitian System}
      In this section, we give examples of bidirectional skin effect in a $2$-dimensional non-Hermitian system with TRS$^\dagger$.
      \par

\begin{figure}
          \centering
              \subfigure[]{\includegraphics[scale=0.45]{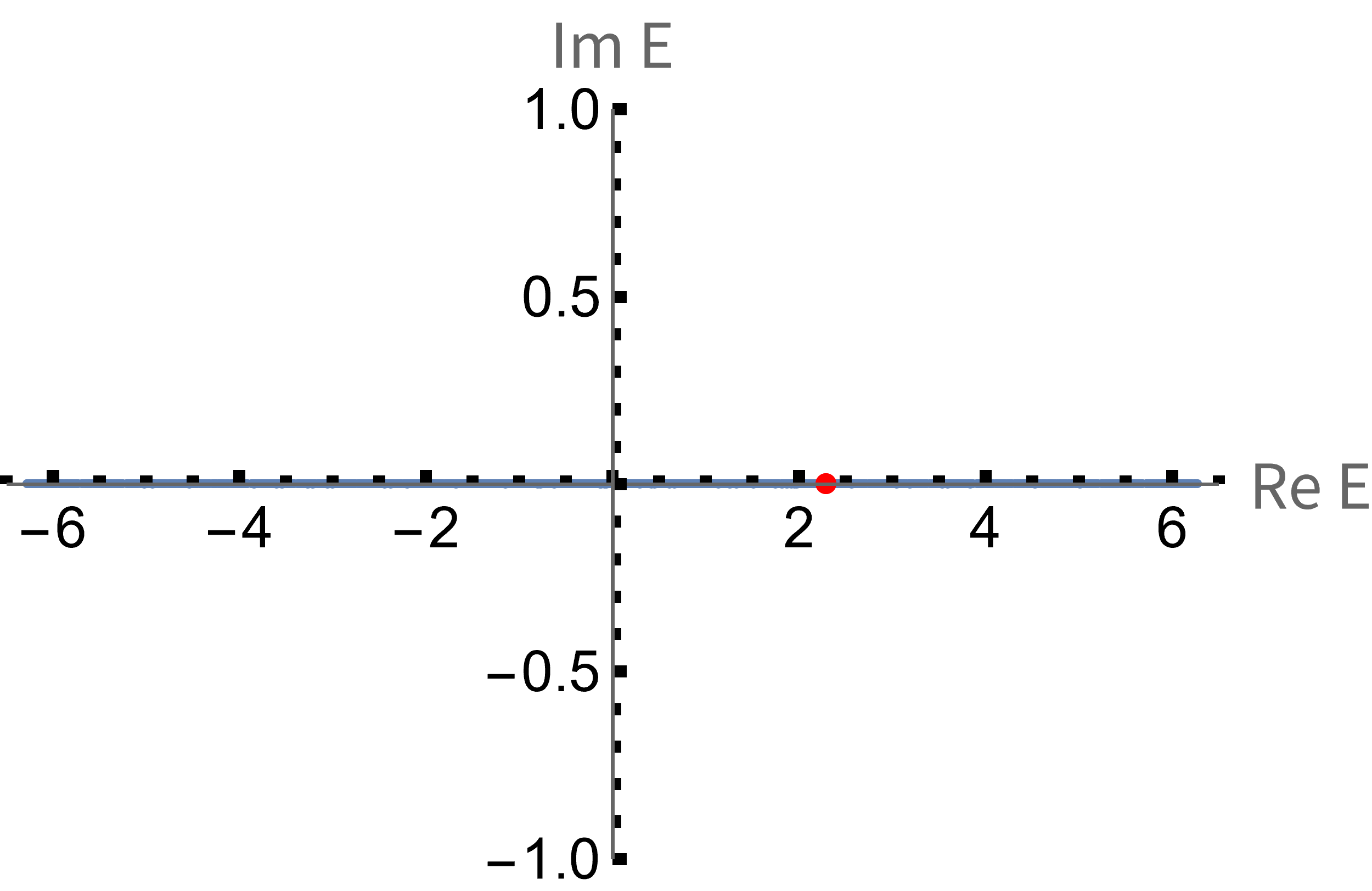} \label{fig6a}}
              \quad
              \subfigure[]{\includegraphics[scale=0.3]{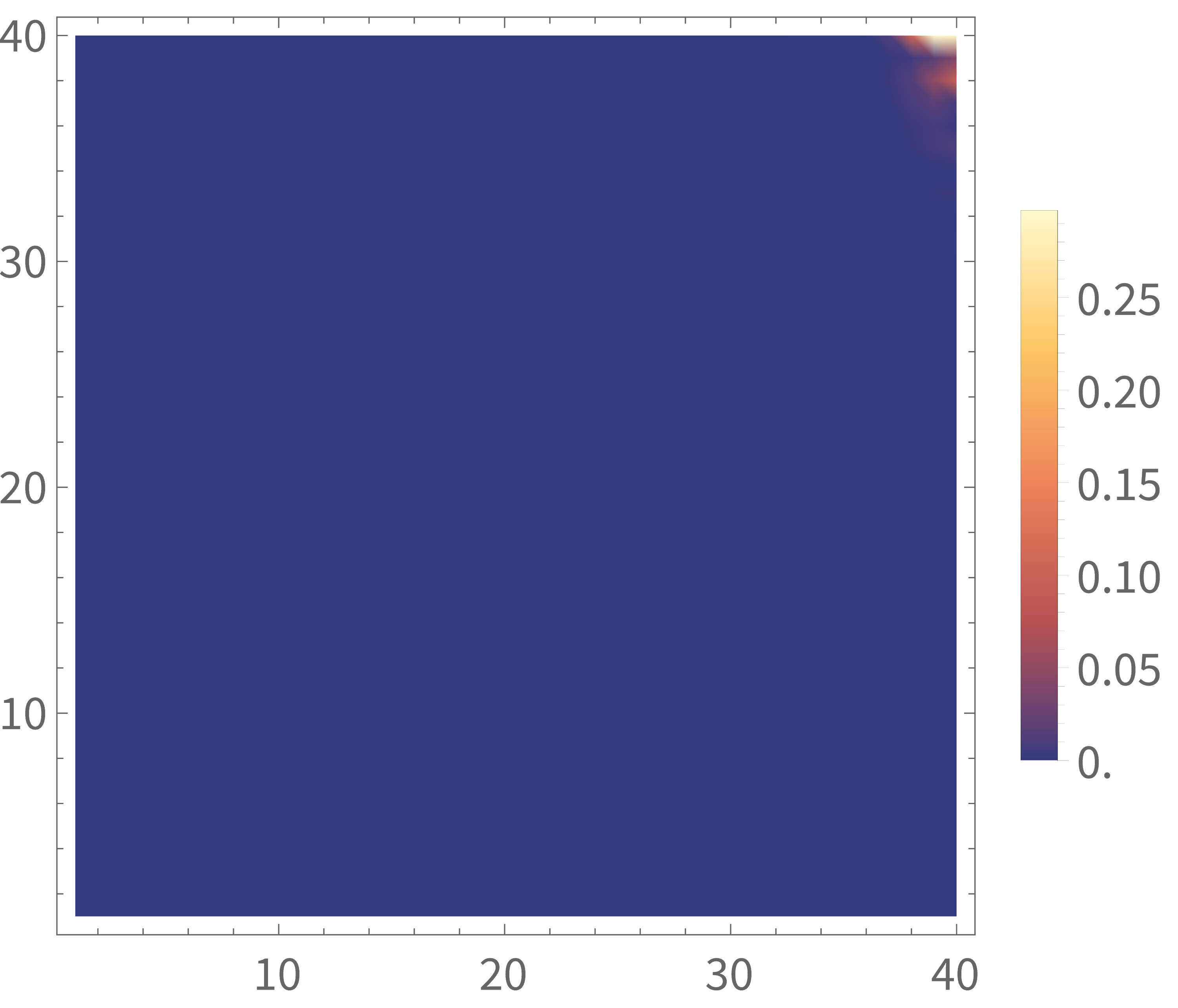} \label{fig6b}}
              \quad
              \subfigure[]{\includegraphics[scale=0.3]{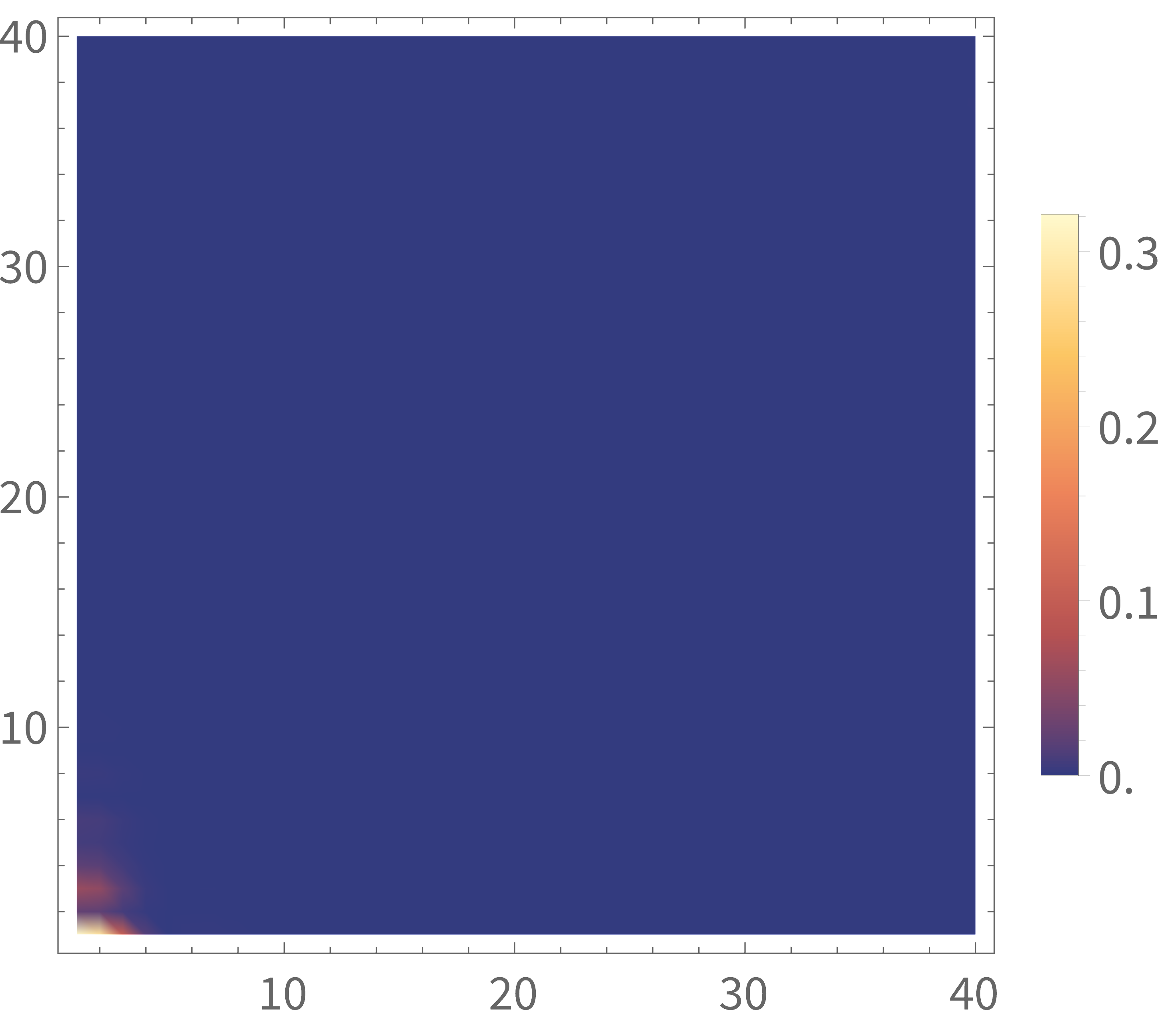} \label{fig6c}}
              \caption{(a)The blue line is the spectrum of the system given by Eq.\eqref{S47} under OBC for the case $\gamma = 0$. The red dot is $E_{1} = 2.29$. (b) and (c) are distributions of two eigenstates degenerate at $E_{1}$. \   The size of the system is $40\times40$.}
              \label{figs6}
        \end{figure}

\begin{figure}
          \centering
          \subfigure[]{\includegraphics[scale=0.45]{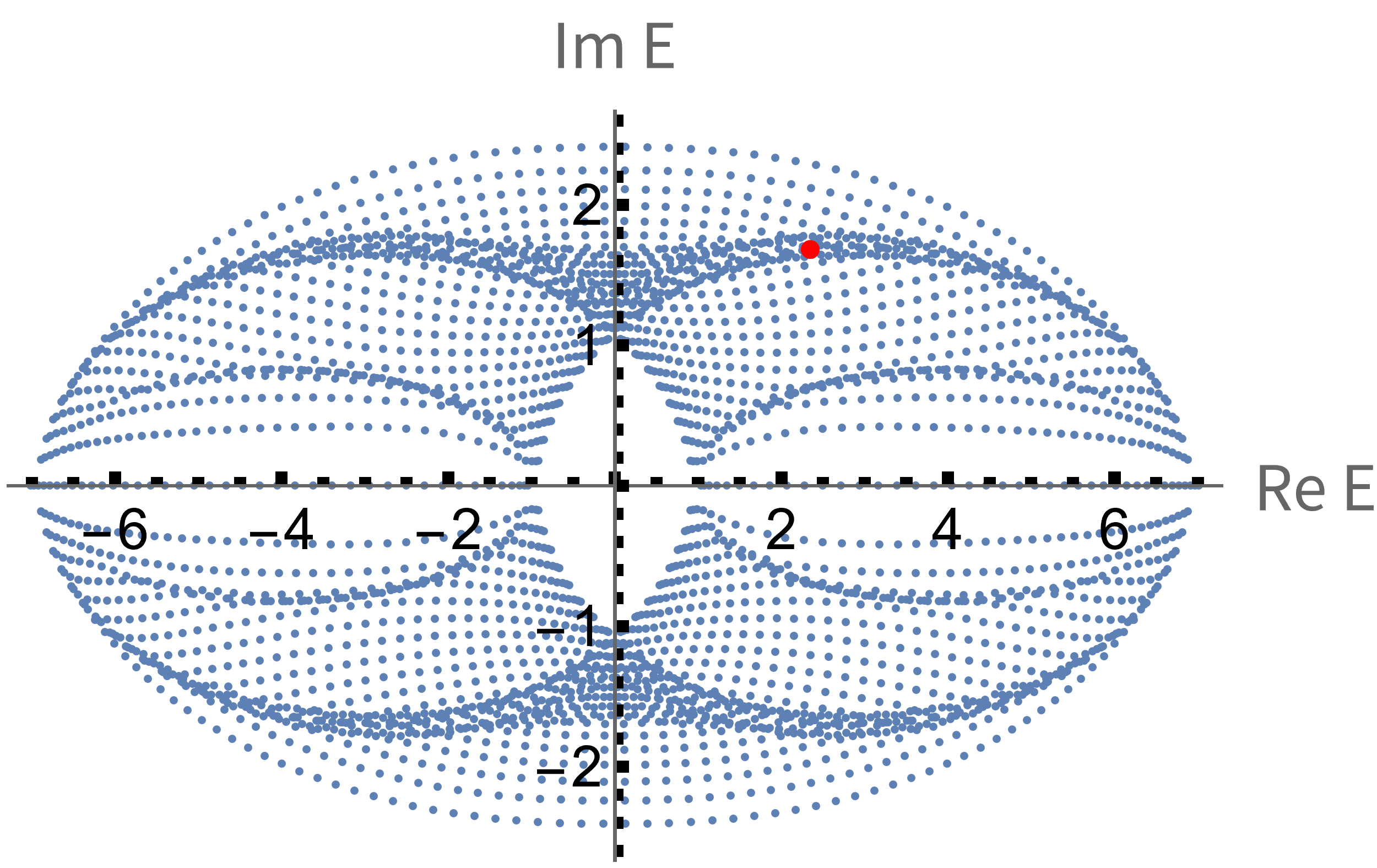} \label{fig7a}}
          \qquad
          \subfigure[]{\includegraphics[scale=0.3]{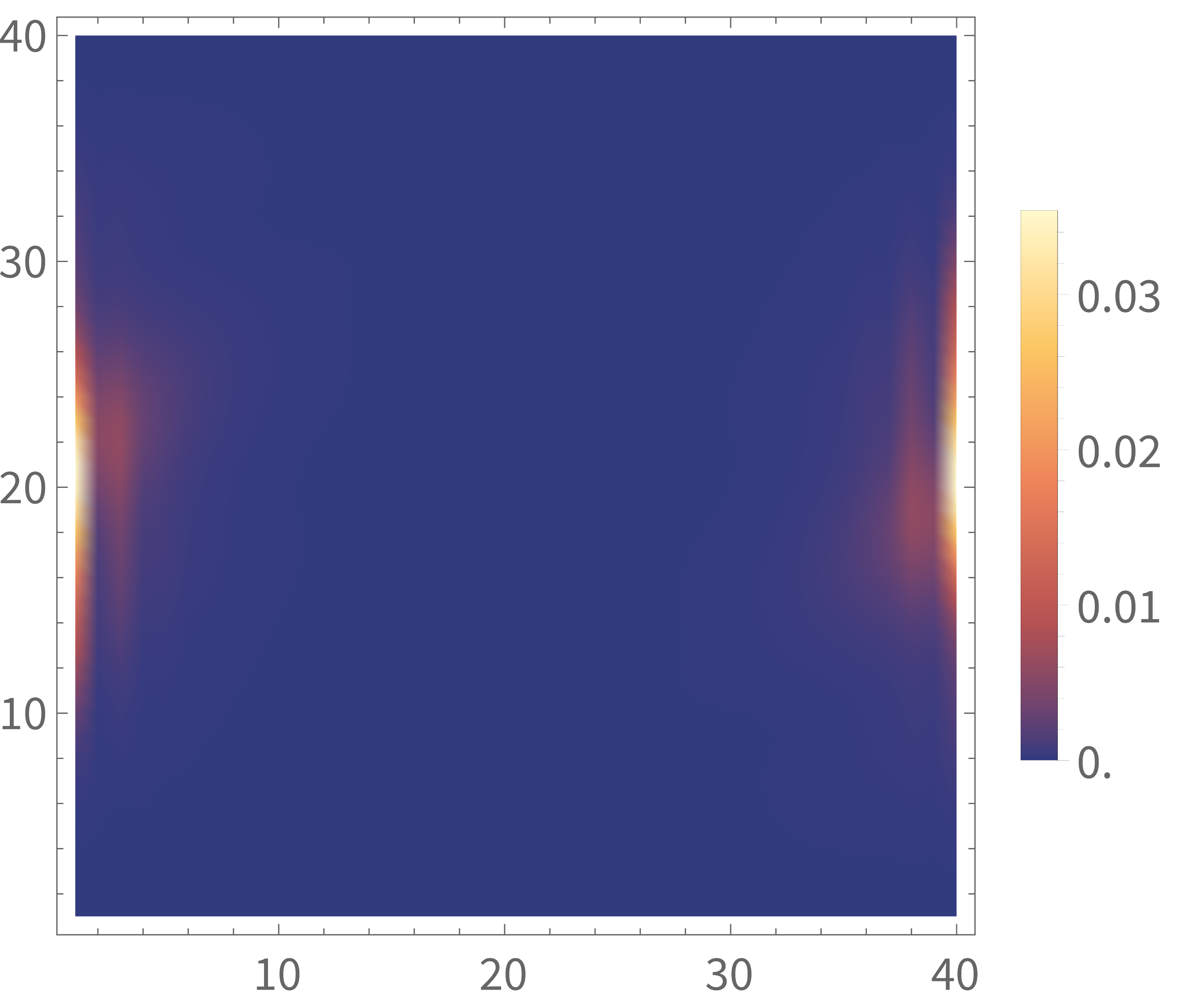} \label{fig7b}}
          
          \caption{(a)The blue dots are the spectrum of the system given by Eq.\eqref{S47} under OBC for the case $\gamma = 0.1$. The red dot is $E_{2} = 2.35 +1.68 i$. (b) is the distribution of eigenstate corresponding to $E_{2}$. \   The size of the system is $40\times40$.}
          \label{figs7}
        \end{figure}

\begin{figure}
          \subfigure[]{\includegraphics[scale=0.45]{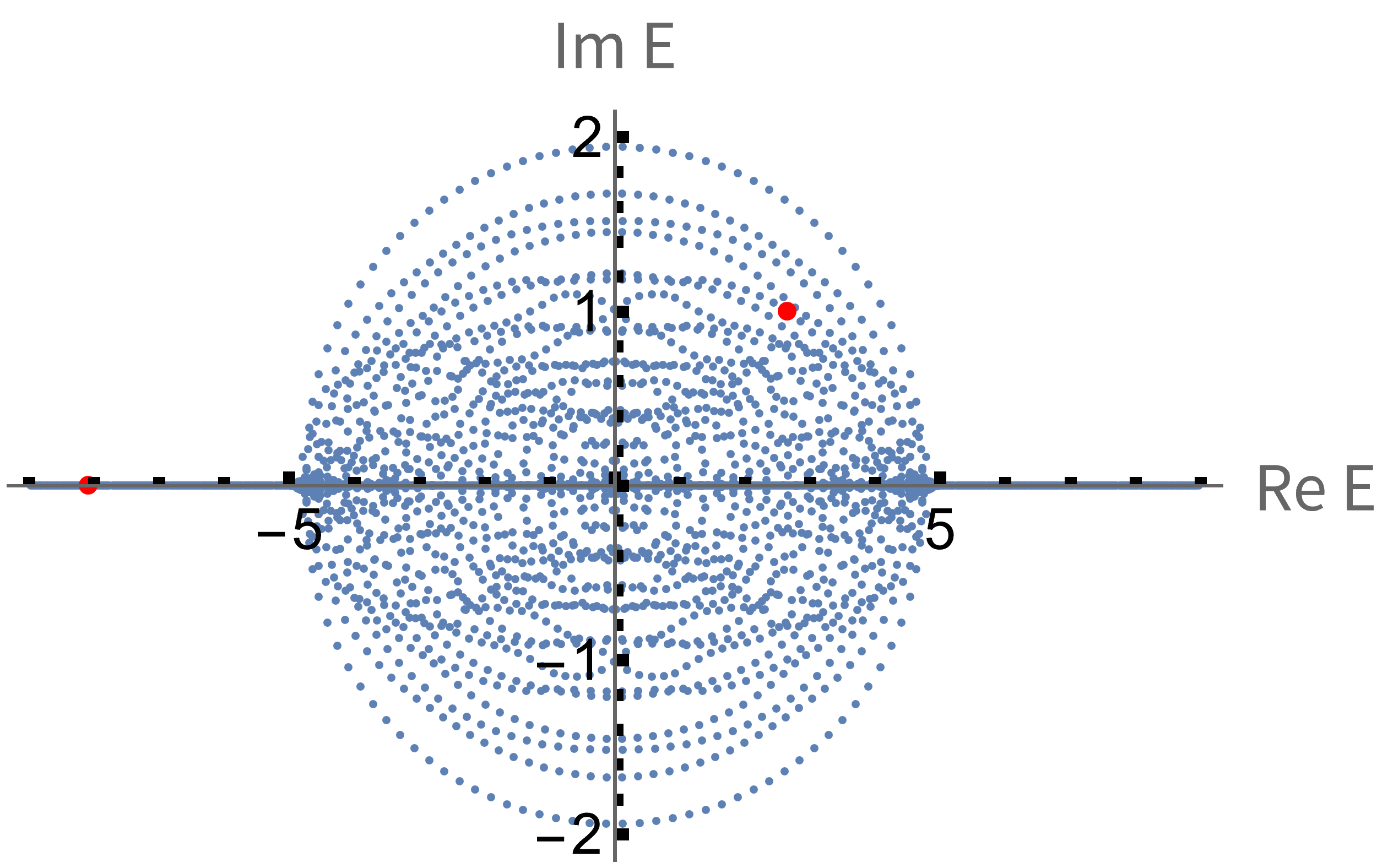} \label{fig8a}}
          \quad
          \subfigure[]{\includegraphics[scale=0.3]{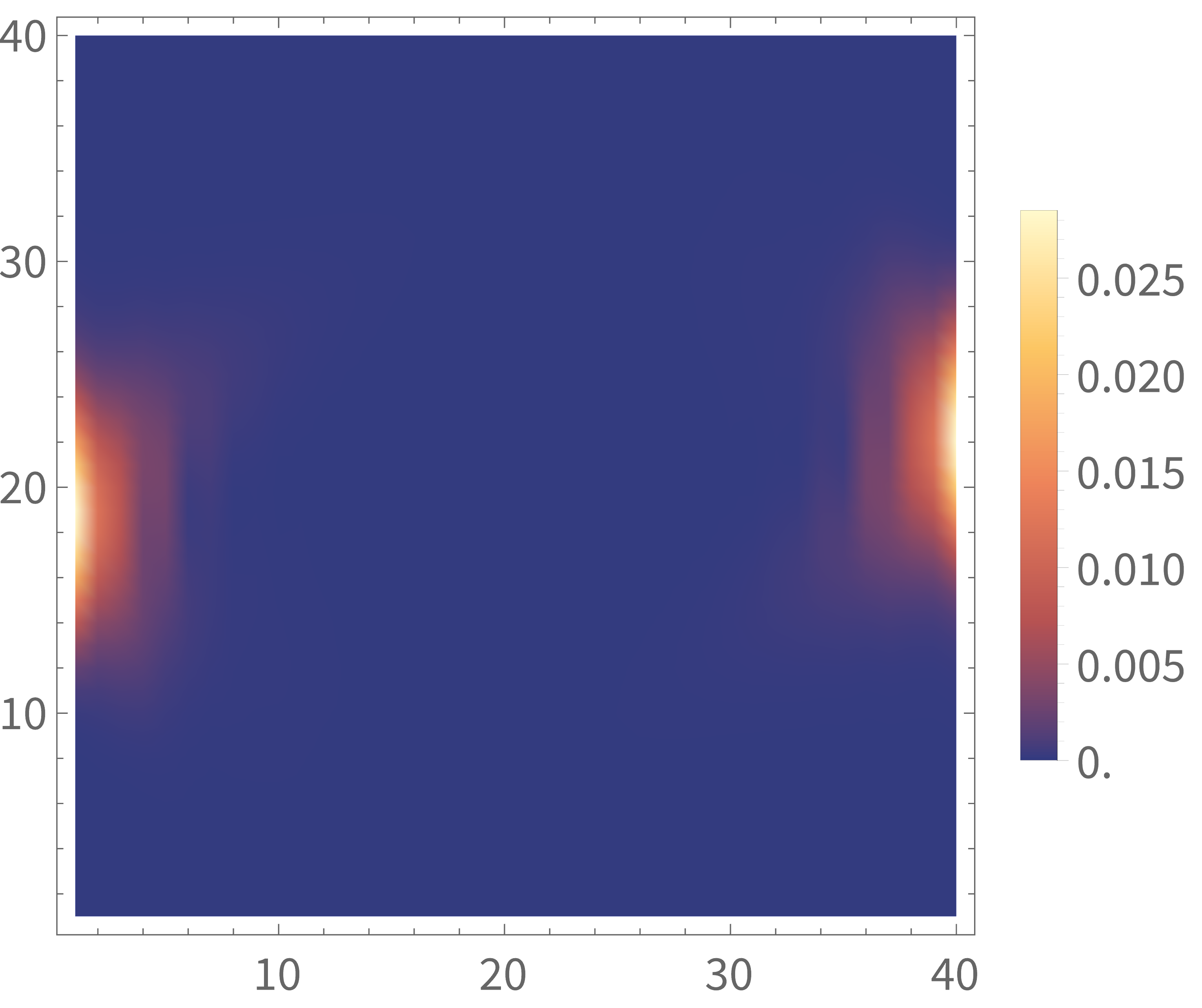} \label{fig8b}}
          \quad
          \subfigure[]{\includegraphics[scale=0.3]{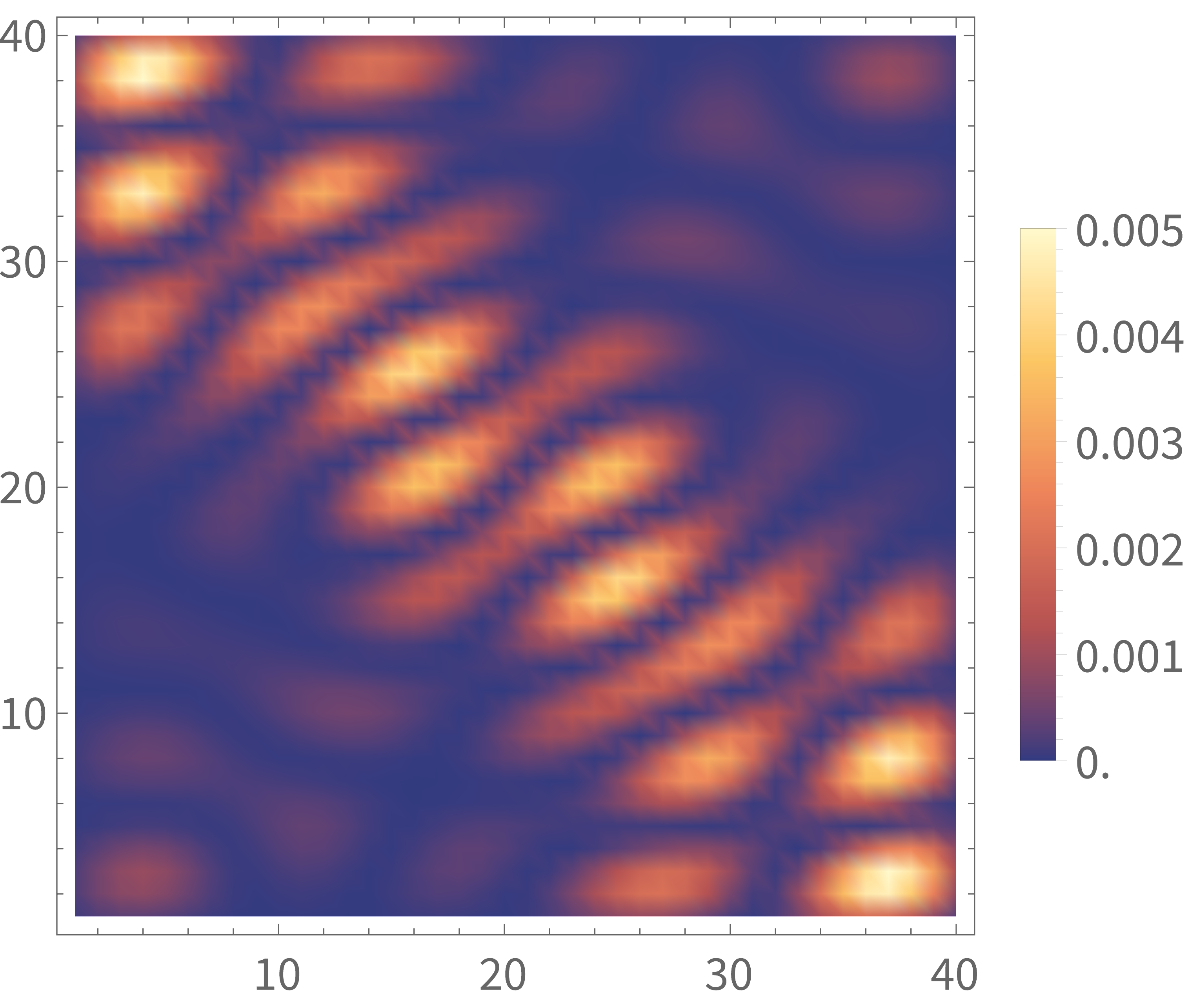} \label{fig8c}}
          \caption{(a)The blue dots are the spectrum of the system given by Eq.\eqref{S47} under OBC for the case $\gamma = 2$. The red dots are $E_{3} = 2.65 + i$ and $E_{4} = -8.09$. (b) and (c) are distributions of two eigenstates corresponding to $E_{3}$ and $E_{4}$ respectively. \   The size of the system is $40\times40$.}
          \label{figs8}
        \end{figure}

      Consider such a model, whose Hamiltonian under PBC is
        \begin{equation}
           H_{TRS^{\dagger}} (k_x,k_y) = 
             \begin{pmatrix}
               t_{-1} e^{-i k_x} + t_1 e^{i k_x} + w_{-1} e^{-i k_y} + w_1 e^{i k_y} &
               \gamma
               \\
               \gamma  &
               t_{-1} e^{i k_x} + t_1 e^{-i k_x} + w_{-1} e^{i k_y} + w_1 e^{-i k_y}
             \end{pmatrix}.
             \tag{S47}
             \label{S47}
        \end{equation}
      The operator of TRS$^{\dagger}$ of this system is 
        \begin{equation}
          \hat{\mathcal{C}}_{+} = U_{\mathcal{C}_{+}} K  = 
            \begin{pmatrix}
              0  &  1  \\
              1  &  0
            \end{pmatrix}
            K.
            \tag{S48}
            \label{S48}
        \end{equation}
      Here, we take $t_1 = 1$, $t_{-1} = 3$, $w_1 = 1$, $w_{-1} = 2$. 
      For $\gamma = 0$, there are two eigenstates degenerated at the same eigenenergy and localized at opposite boundary~(Fig.\ref{figs6}).   
      For $\gamma = 0.1$, the degeneration disappears and bidirectional skin modes appear in the system~(Fig.\ref{figs7}).
      Then, we consider the case, $\gamma = 2$. In this case, the system has bidirectional skin mode and extended mode simultaneously~(Fig.\ref{figs8}).

      \par
      
\begin{figure}
          \subfigure[]{\includegraphics[scale=0.4]{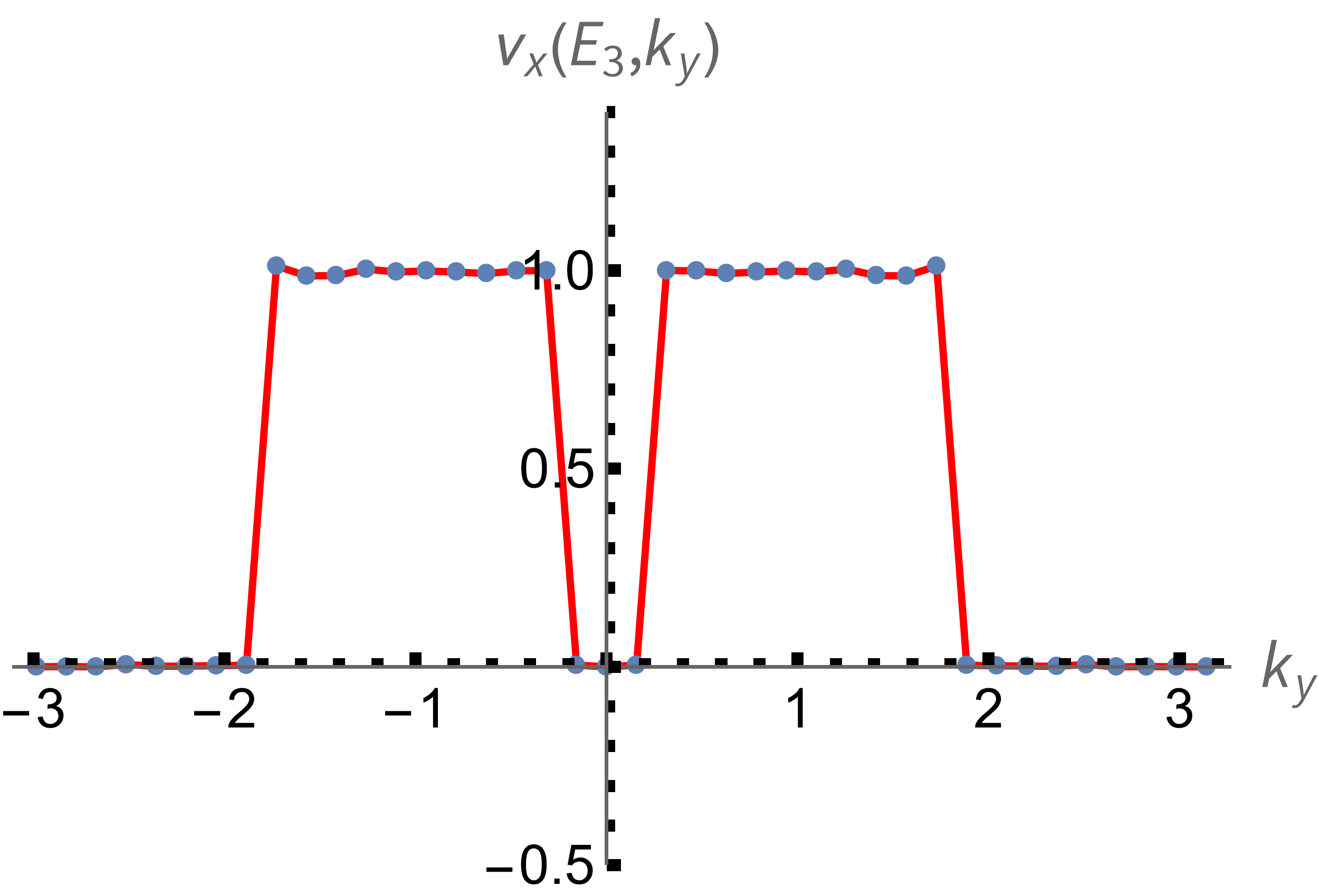} \label{fig9a}}
          \qquad
          \subfigure[]{\includegraphics[scale=0.4]{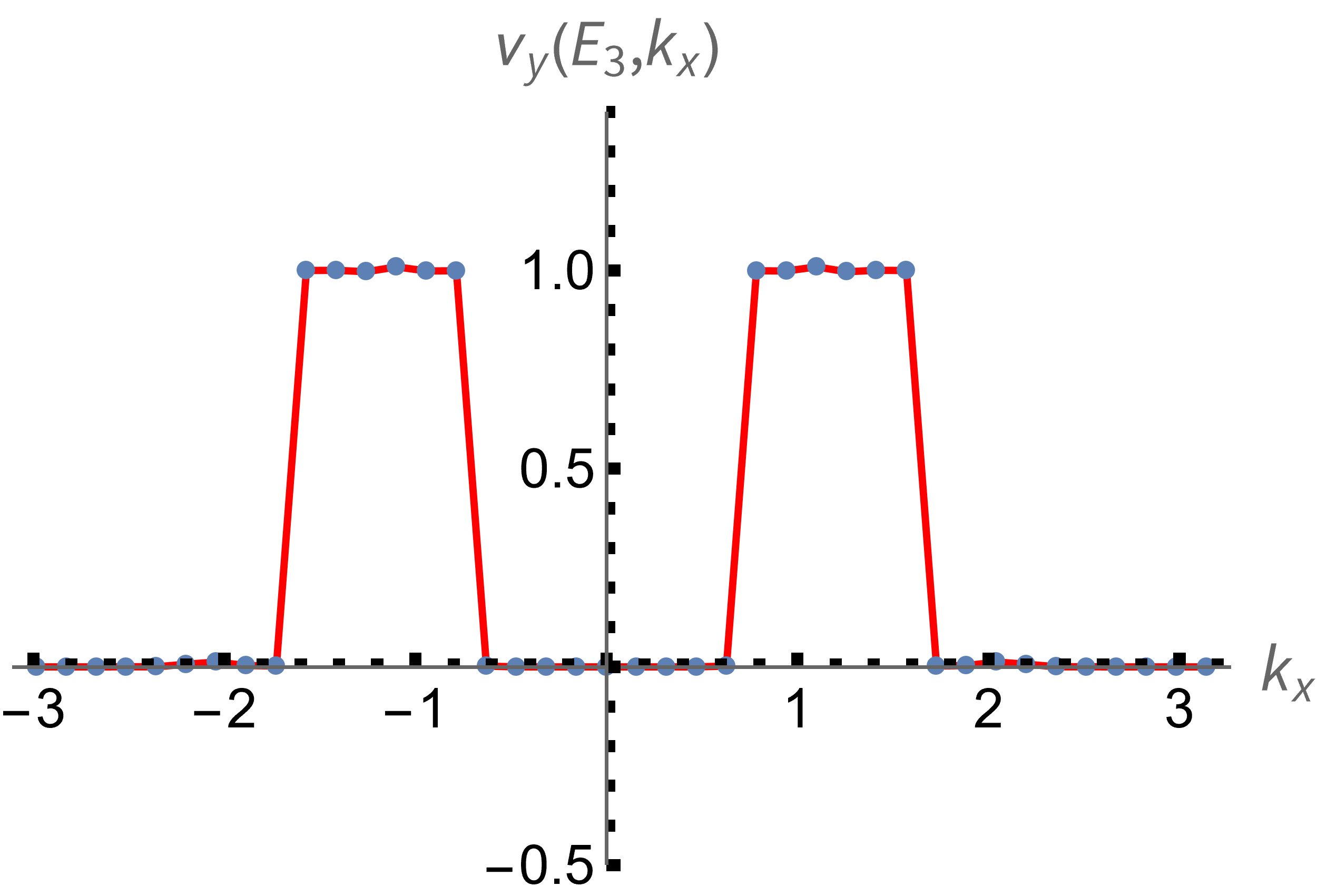} \label{fig9b}}
          \qquad
          \subfigure[]{\includegraphics[scale=0.4]{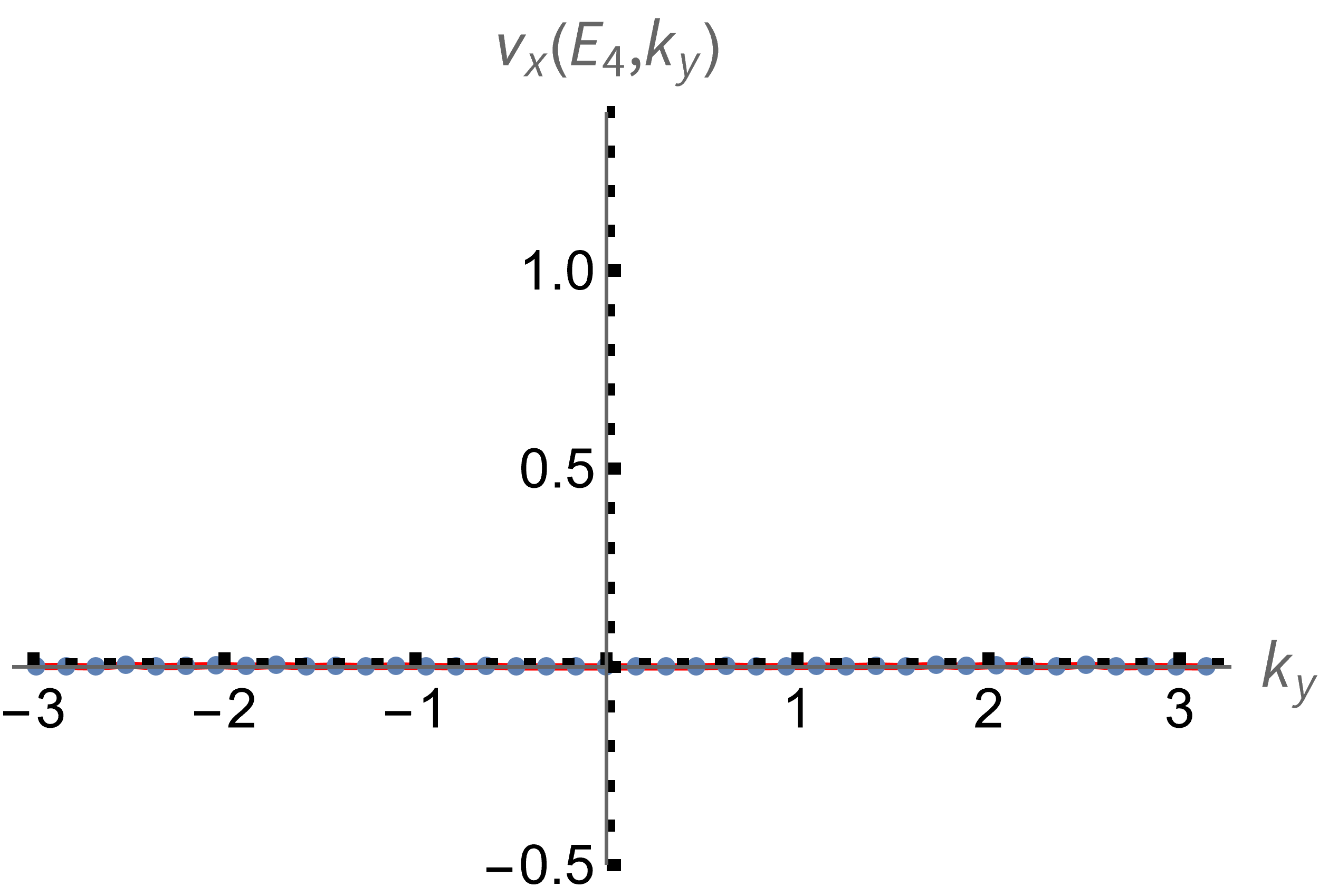} \label{fig9c}}
          \qquad
          \subfigure[]{\includegraphics[scale=0.4]{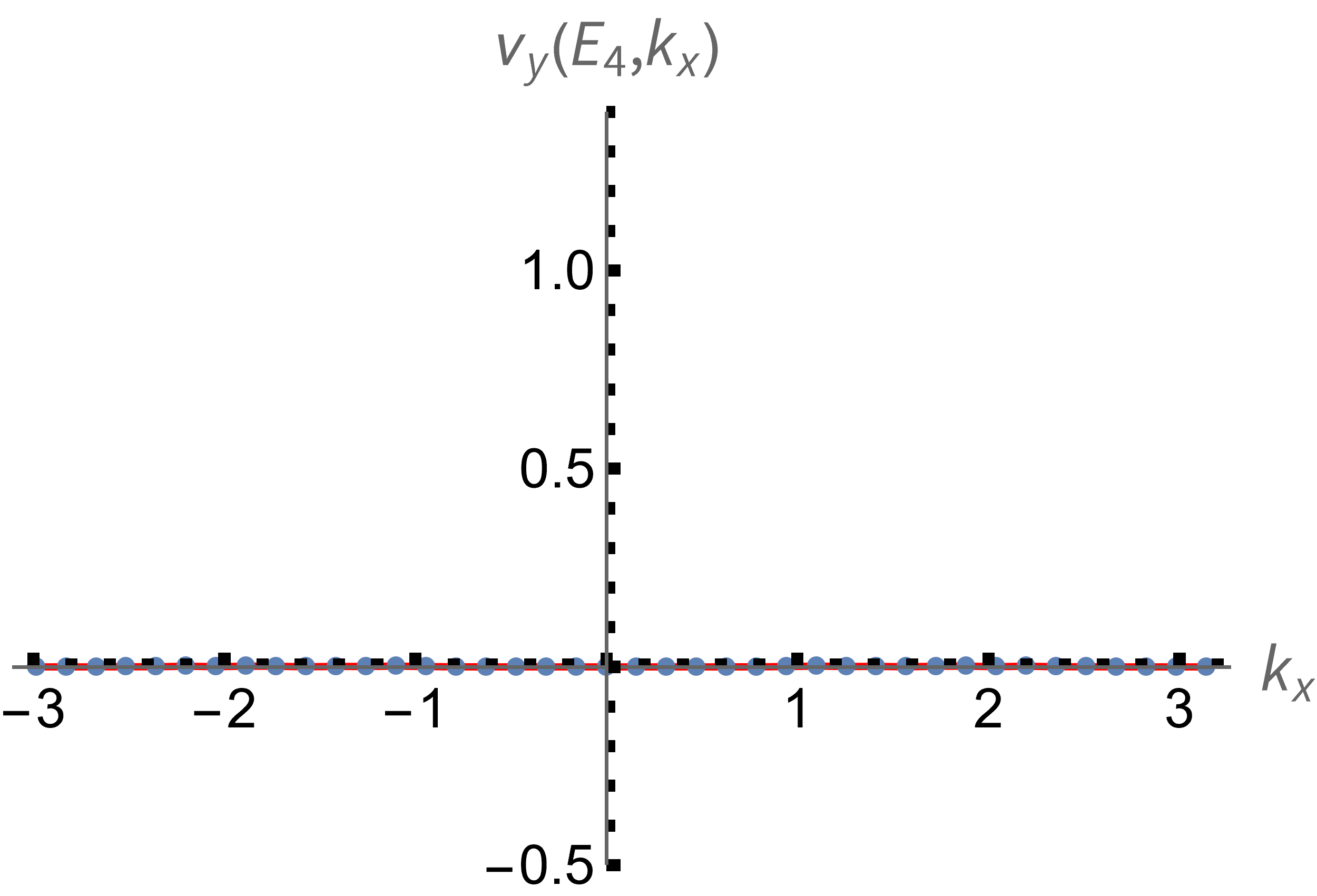} \label{fig9d}}
          \caption{The blue dots in (a), (b), (c) and (d) are values of $\nu_x (E_3,k_y)$, $\nu_y (E_3,k_x)$, $\nu_x (E_4,k_y)$ and $\nu_y (E_4,k_x)$ respectively for $k_x \in [-\pi, \pi]$ or $k_y \in [-\pi, \pi]$ in the case that $\gamma = 2$. }
          \label{figs9}
        \end{figure}

      The two bands of $ H_{TRS^{\dagger}} (k_x,k_y)$ are
        \begin{equation}
           E_{\pm} (k_x , k_y) = (t_1 + t_{-1}) \cos(k_x) + (w_1 + w_{-1}) \cos(k_y) \pm 
           \sqrt{\gamma^2 - \left[ (t_1 - t_{-1}) \sin(k_x) + (w_1 - w_{-1}) \sin(k_y) \right]^2}.
           \tag{S49}
           \label{S49}
        \end{equation}
      $E_{\pm} (k_x , k_y)$ are related by TRS$^\dagger$. Henc, according to the main text, the winding numbers of the two bands about $E$ satisfy
        \begin{equation}
          w_x^{(+)} (E, k_y) = - w_x^{(-)} (E, k_y) \qquad
          w_y^{(+)} (E, k_x) = - w_y^{(-)} (E, k_x).
          \tag{S50}
          \label{S50}
        \end{equation} 
      Then, according to Eq.(29) in the main text,
        \begin{equation}
          \nu_x (E,k_y) = |w_x^{(+)} (E, k_y)|  
          \qquad
          \nu_y (E,k_x) = |w_y^{(+)} (E, k_x)|.
          \tag{S51}
          \label{S51}
        \end{equation}
      For $\gamma = 2$, values of the topological invariants, $\nu_x (E,k_y)$ and $\nu_y (E,k_x)$ for $E = E_3$ and $E = E_4$ are given in Fig.\ref{figs9}. 
      For $E_3$, the eigenstate corresponding to it is bidirectional skin mode and $\nu_x (E_3,k_y)$ and $\nu_y (E_3,k_x)$ can be non-zero for some values of $k_x$ and $k_y$.
       For $E_4$, the eigenstate corresponding to it is extended and $\nu_x (E_4,k_y)= \nu_y (E_4,k_x) = 0$ for all $k_x, k_y \in [-\pi, \pi]$. This is consistent with our conclusion in the main text.


\end{document}